\documentclass[a4paper]{article}

\RequirePackage[OT1]{fontenc}
\RequirePackage{amsthm,amsmath}
\RequirePackage[numbers]{natbib}
\RequirePackage[colorlinks,citecolor=blue,urlcolor=blue]{hyperref}
\usepackage{latexsym,graphicx,subcaption,verbatim,nameref}
\usepackage{diagbox}
\usepackage[all,cmtip]{xy}
\usepackage[toc,title,titletoc]{appendix}
\usepackage[capposition=top]{floatrow}
\usepackage{rotating}
\usepackage{threeparttable,booktabs}
\usepackage{csquotes} 
\usepackage{chngcntr}
\usepackage{subcaption}
\usepackage[affil-it]{authblk}
\usepackage{graphicx}
\usepackage{esint}
\usepackage{relsize}
\usepackage{color}

\counterwithin{figure}{section}
\counterwithin{table}{section}
\numberwithin{equation}{section}
\theoremstyle{plain}
\newtheorem{theorem}{Theorem}[section]
\newtheorem{property}[theorem]{Property}
\newtheorem{proposition}[theorem]{Proposition}

\theoremstyle{definition}
\newtheorem{definition}[theorem]{Definition}
\newtheorem{lemma}[theorem]{Lemma}
\newtheorem{principle}[theorem]{Principle}
\newcommand*{\Scale}[2][4]{\scalebox{#1}{$#2$}}%
\renewcommand{\footnotesize}{\fontsize{8.5pt}{10pt}\selectfont}

\begin{document}

\title{Stochastic Process Model and Health Data: A Full Maximum Likelihood Method to Hospital Charge and Length of Stay Data }

\author{Xiaoqi Zhang%
\thanks{Email: \texttt{xiaoqizh@buffal.edu.edu}; Corresponding Author.}}
\affil{Department of Mathematics,\\ State University of New York at Buffalo}

\author{John Ringland%
\thanks{Email: \texttt{ringland@buffal.edu.edu.}}}
\affil{Department of Mathematics,\\ State University of New York at Buffalo}

\maketitle
%
%
%
%
%
%
%

\begin{abstract}
We extend the model used in \cite{gardiner2002longitudinal} and \cite{polverejan2003estimating} through deriving an explicit expression for the joint probability density function of hospital charge and length of stay (LOS) under a general class of conditions. Using this joint density function, we can apply the full maximum likelihood method (FML) to estimate the effect of covariates on charge and LOS. By FML, the endogeneity issues arisen from the dependence between charge and LOS can be efficiently resolved. As an illustrative example, we apply our method to real charge and LOS data sampled from New York State's Statewide Planning and Research Cooperative System 2013 (SPARCS 2013). We compare our fitting result with the fitting to the marginal LOS data generated by the widely used Phase-Type model, and conclude that our method is more efficient in fitting.  
\end{abstract}

%
\renewcommand{\footnotesize}{\fontsize{5pt}{6pt}\selectfont}
\section{Introduction}

Rising expenditures and constraints on health care budgets have prompted the development of a variety of methods for the analyses of hospital charge and length of stay (LOS) as discussed in \cite{gold1996panel}, 
\cite{lipscomb1998predicting} and
\cite{lin1997estimating}. Identifying the important determinants of charge and LOS is a critical step in optimizing the allocation of healthcare resources. However, it is always challenging to find regression models that can efficiently resolve the potential endogenous issues caused by the complicated causal relationship 
among charge, LOS and many other factors. In fact, total charge is naturally increasing as the increasing of LOS because more healthcare resources would be consumed during a longer stay. On the other hand, LOS might also be influenced by the accumulation of total charge over time because charge reflects the potential economic incentives to both of patient's and doctor's discharge decision. Finally, factors, like patient's diagnosis, severity of illness, insurance status and so on, could affect the interaction between the charge accumulation and discharge decision (or LOS). 

One efficient solution to endogenous issues is to model the charge, discharge decision (or LOS) and the interaction between them \enquote{internally} so that a parametric joint distribution of the charge and LOS can be derived explicitly from the model. All external factors come into affecting charge and LOS only through those parameters involved in the joint distribution. 

Following this direction, \cite{gardiner2002longitudinal} and  \cite{polverejan2003estimating} estimated a model in which both of the charge and LOS were included endogenously. However, neither of them attempted to derive a general expression of the joint density of charge and LOS. Instead, they relied on a few special classes of parametric density functions. As a consequence, \cite{gardiner2002longitudinal} and \cite{polverejan2003estimating} cannot test the goodness of fitting of their model to the joint density of charge and LOS, meanwhile they have to face the technical challenge of how to select viable instrument variables.

In this paper, we shall extend the model introduced in \cite{gardiner2002longitudinal} and  \cite{polverejan2003estimating} in the direction that we will derive an explicit expression of the joint probability density of charge and LOS. With the joint density function, we can use the full maximum likelihood method (FML) to estimate the parameters in contrast to introduce instrument variables as did in \cite{gardiner2002longitudinal}. The advantage of FML is that the resulting estimators are asymptotically unbiased, efficient (endogeneity resolved) and the asymptotic distribution of estimators is known to be normal.   

In section 2, we re-introduce the model discussed in \cite{gardiner2002longitudinal} and \cite{polverejan2003estimating}, and derive the marginal and joint distributions of charge and LOS associated with this model. Since the model in consideration is a continuous-time Random Growth Process with a Radom Stopping Time, we will call it RGRST model in short. In section 3, we carry out the regression analysis based on the RGRST model and use FML to do estimation with respect to the charge and LOS data sampled from SPARCS 2013.

\section{Continuous-Time RGRST Process }

\begin{definition}
A continuous time RGRST process is defined as

\begin{equation} \label{defn1}
Y_{t}\left(\omega\right)=Y_{0}\left(\omega\right)+
\int_{0}^{t}I\left(\omega,Y_{s},s\right)
\epsilon_{s}\left(\omega\right)ds
\end{equation}

where $I\left(.,Y_{s},s\right):\Omega\rightarrow\left\{ 0,1\right\} $
is decision variable representing patient's discharge decision, whether
or not to stay in hospital for longer time at each time point $s$, it
takes value 1 if patients decide to stay and 0 otherwise. $\left\{ \epsilon_{t}\right\} $
is a non-negative process characterizing the potential increment of
charge per unit time provided that patient decides to stay.

It is natural to require the decision process $I$ satisfies the following Non-Increasing Property:

\[
s\leq s'\Longrightarrow I\left(\omega,Y_{s}\left(\omega\right),s\right)\geq I\left(\omega,Y_{s'}\left(\omega\right),s'\right)\; a.s.
\]

\end{definition}
The Non-Increasing Property means once if a patient (represented as state
$\omega$) decides to leave hospital at time $s$ ($I_{s}=0$) , he/she
could never choose to come back at later time $s'\geq s$ ($I_{s'}=1$).
Based on the non-increasing property and continuity property, time in
RGRST process can be naturally interpreted as the length of a patient
staying in hospital. Consequently, the random time $T\left(\omega\right):=\min\left\{ t\geq0:\, I\left(\omega,Y_{t}\left(\omega\right),t\right)=0\right\} $
is exactly the length of stay associated with patient $\omega$ and
the random variable $Y_{T}$ represents the total charge at discharge
day.

One important advantage of the continuous time RGRST process is that
both of the marginal and joint distributions induced by $Y_{T}$, $T$
can be solved for explicitly. In the next section, we focus
on the explicit expression of these distributions. 

\subsection{Distributions of Charge and LOS} \label{sec 2.2}

We need the following two important conditional expectations for the derivation of the distributions in interest:

\begin{equation} \label{unique eq}
\begin{aligned}
\tilde{q}\left(y,t\right)=&
E\left(\epsilon_{t}\middle|G_t=y\right)\\
\tilde{q}_{1}\left(y,t\right)=&
E\left(I\left(.,Y_{t},t\right)\middle|G_t=y\right)
\end{aligned}
\end{equation}

where $G_t
:=Y_0+\int_{0}^{t}\epsilon_{s}ds$ is defined to be the potential growth process of charge. In addition, we denote $P\left(.,.\right)$ as the joint probability density function induced by charge and LOS; denote $p\left(.,t\right)$ and $\tilde{p}\left(.,t\right)$ as the probability density with respect to $Y_t$ and $G_t$.

Using the following two properties that we require the functions $\tilde{q}$, $\tilde{q}_1$ and $p\left(.,0\right)$ (the initial probability density function associated with $Y_0$) to satisfy:

\begin{property}\label{Property 1}
$\tilde{q}$, $\tilde{q}_{1}$, and $p\left(.,0\right)$ satisfy that $0\leq\tilde{q}_{1}\leq1$, $\tilde{q}\geq0$ and $p\left(.,0\right)$
is a well-defined probability density function.
\end{property}

\begin{property}\label{Property 3}
The directional derivative for the function $\tilde{q}_{1}$
in the direction $\left(\tilde{q},1\right)$ is non-positive
for all $\left(y,t\right)$.
\end{property}
We could prove our main result, Theorem \ref{existence and uniqueness}.

\begin{theorem}
\label{existence and uniqueness}
Given functions $\tilde{q}$, $\tilde{q}_{1}$ and $p\left(.,0\right)$ and suppose all of them are $C^{2}\left([0,\infty)^{2}\right)$ functions. 

Then there exists a RGRST process $\left\{Y_t:\,t\in[0,\infty)\right\}$ satisfying Equation \ref{unique eq} 
if and only if $\tilde{q}$, $\tilde{q}_{1}$ and $p\left(.,0\right)$ satisfy Property \ref{Property 1} and \ref{Property 3}.

Moreover, the process $\left\{Y_t:\,t\in[0,\infty)\right\}$ satisfying Equation \ref{unique eq} is unique in the sense that any two processes satisfying Equation \ref{unique eq} would induce the same time-dependent density function $p\left(y,t\right)$ and the same joint distribution of charge and LOS.
\end{theorem}

Theorem \ref{existence and uniqueness} follows from the following two lemmas:
\begin{lemma}\label{lemma5}
Given a RGRST process $\left\{ X_{t}:t\in[0,\infty)\right\} $ as
in definition \ref{defn1} with the initial density $p\left(.,0\right)$. Suppose the corresponding functions $\tilde{q}$ and $\tilde{q}_{1}$ as defined in Equation \ref{unique eq}
satisfy the requirement in Theorem \ref{existence and uniqueness}, then  the process $\left\{ X_{t}\right\}$ is equivalent to the following RGRST process defined on the
probability space $(\Omega:=\left[0,1\right]\times
[0,\infty),\mathcal{B},dm\times p\left(.,0\right)dm)$

\begin{equation}\label{q1_tilde construction}
Y_{t}\left(\omega,y_{0}\right)=y_{0}+\int_{0}^{t}\tilde{q}\left(Y_{s},s\right)\cdot\mathbf{1}_{\left\{ \omega\leq\tilde{q}_{1}\left(\tilde{g}^{-1}\left(y_{0},0,s\right),s\right)\right\} }ds
\end{equation}
in the sense that both RGRST processes have the same time-dependent density function $p\left(y,t\right)$ and the same joint distribution of charge and LOS. Moreover, the functions $\tilde{q}$
and $\tilde{q}_{1}$ involved in above constructions satisfy

\[
-\frac{\partial\tilde{q}_{1}}{\partial y}\cdot\tilde{q}-\frac{\partial\tilde{q}_{1}}{\partial t}\geq0.
\]
where $\tilde{g}\left(y,t,s\right)$, viewed as a family of functions in
variable $s$, solves the initial value problems 

\begin{equation}
\label{ode2}
\begin{aligned}
\frac{dy}{dt} =  -\tilde{q}\left(y,t-s\right)\\
\tilde{g}\left(y,t,0\right) =  y
\end{aligned}
\end{equation}
and $\tilde{g}^{-1}$ is defined as the inverse to $\tilde{g}$ such
that $\tilde{g}^{-1}\left(y,0,t\right):=\left\{ x:\tilde{g}\left(x,t,t\right)=y\right\}$.
\end{lemma}

\begin{lemma}\label{lemma6}
Given a probability density function $p\left(.,0\right)$ and functions
$\tilde{q},\tilde{q}_{1}:[0,\infty)^{2}\rightarrow[0,\infty)$ such
that Property \ref{Property 3} holds for $\tilde{q}$ and $\tilde{q}_{1}$,  then there exist RGRST process $\left\{ Y_{t}:t\in[0,\infty)\right\} $
which can be constructed in the same way as in Equation \ref{q1_tilde construction} and the process
$\left\{ Y_{t}:t\in[0,\infty)\right\} $ has the time-dependent charge
density 
\begin{equation} \label{time dependent density 2}
p\left(y,t\right)=\int_{0}^{t}-\left(\frac{\partial\tilde{q}_{1}}{\partial y}\cdot\tilde{q}+\frac{\partial\tilde{q}_{1}}{\partial t}\right)\left(y,s\right)\cdot\tilde{p}\left(y,s\right)ds+\tilde{q}_{1}\left(y,t\right)\cdot\tilde{p}\left(y,t\right)
\end{equation}

Moreover, the joint distribution of final charge at discharge day $Y_T$ and the LOS $T$ can be given as follows:

\begin{equation} \label{joint density 2nd expression}
P\left(y,t\right)=\tilde{p}\left(y,t\right)\cdot\left(-\frac{\partial\tilde{q}_{1}}{\partial y}\cdot\tilde{q}-\frac{\partial\tilde{q}_{1}}{\partial t}\right)\left(y,t\right)
\end{equation}
where $\tilde{p}\left(y,s\right):=\frac{\partial\tilde{g}\left(y,s,s\right)}{\partial y}\cdot p\left(\tilde{g}\left(y,s,s\right),
0\right)$ and the function $\tilde{g}$ is constructed from $\tilde{q}$ in
the same way as in Lemma \ref{lemma5}.
\end{lemma}

Proof for the two lemmas is in Appendix A. Lemma \ref{lemma5} gives us a way to characterize a RGRST process
through specifying the conditional probability of $\left\{ LOS>t\right\} $
on the deterministic version of the potential growth process. Lemma \ref{lemma6} provides an explicit description (Equation \ref{joint density 2nd expression}) of the joint distribution of final charge ($Y_T$) and the LOS ($T$). Based on the two lemmas,
we can simply sketch the proof for Theorem \ref{existence and uniqueness}:
\begin{proof}
Necessity of Property \ref{Property 1} and \ref{Property 3}.

Property \ref{Property 1} just says $p\left(.,0\right)$, $q$ and $q_{1}$ are well-defined
probability density and/or conditional expectations of some non-negative
random variables. Therefore, for an arbitrary RGRST process $\left\{ X_{t}\right\} $, Property
\ref{Property 1} always hold. Lemma \ref{lemma5} implies Property \ref{Property 3}. 

Sufficiency of Property \ref{Property 1} and \ref{Property 3} is directly given by Lemma \ref{lemma6}.
\end{proof}

\subsubsection{Application of Equation \ref{joint density 2nd expression}}
The expression \ref{joint density 2nd expression} is the key to resolve the endogeneity brought by the dependence between charge and LOS.

In fact, the endogeneity in the setting of charge and LOS comes from the following structural equations:

\begin{equation}\label{structural equation}
\begin{aligned}
\ln Y_T =& \beta_{y,0}+\beta_{y}\cdot X + f\left(T\right) + \epsilon_{Y_T}\\
\ln T = & \beta_{t,0}+\beta_{t}\cdot X' + g\left(Y_T\right) + \epsilon_{LOS},
\end{aligned}
\end{equation}
Equation \ref{structural equation} describes how the charge ($Y_T$) and LOS ($T$) are dependent on each other (through function $f$ and $g$) and on explanatory variables $X$ and $X'$. It is widely known that endogeneity exists when estimate such structural equations as \ref{structural equation} and it causes the OLS estimators biased and inconsistent. 

To overcome the failure of OLS estimation when endogeneity exists, we use Full Maximum Likelihood method (FML) to replace OLS. As known, FML can always generate consistent and normally distributed estimation of parameters under a general class of conditions. 

In the setting of estimating Equation \ref{structural equation}, given an explicit and parametric expression of the joint probability density function of charge and LOS conditional on each given profile of explanatory variables (denoted as $P_{\alpha}(\,.\,,\,.\,|\beta_{y,0}+\beta_{y}\cdot X,\,\beta_{t,0}+\beta_{t}\cdot X')$ with the subscript $\alpha$ representing the vector of parameters that describe the functional form of $P$), the FML method guarantees that the resulting estimator $\vec{v}_n:=\left(\hat{\alpha},\,\hat{\beta}_y,\, \hat{\beta}_t,\,\hat{\beta}_{y,0},\,\hat{\beta}_{t,0}\right)$ is consistent and asymptotically normally distributed, where the subscript $n$ indicates that the estimator is generated from an i.i.d. sample with size $n$. 

Moreover, using the FML estimator $\vec{v}_n$, we can express the estimators for the function $f$ and $g$ as below:
\begin{equation}\label{estimators}
\begin{aligned}
\hat{f}_n\left(T\right)=
E_{P_{\hat{\alpha}}\left(.,\,T|\hat{\beta}_{y,0}+\hat{\beta}_{y}\cdot X,\hat{\beta}_{t,0}+\hat{\beta}_{t}\cdot X'\right)}\left(Y_T\right)- \hat{\beta}_{y,0}-\hat{\beta}_{y}\cdot X\\
\hat{g}_n\left(Y_T\right)=
E_{P_{\hat{\alpha}}\left(Y_T,\, .|\hat{\beta}_{y,0}+\hat{\beta}_{y}\cdot X,\hat{\beta}_{t,0}+\hat{\beta}_{t}\cdot X'\right)}\left(T\right)- \hat{\beta}_{t,0}-\hat{\beta}_{t}\cdot X',
\end{aligned}
\end{equation}

Given the consistency and asymptotically normal distribution of $\vec{v}_n$, we can verify that the estimators in \ref{estimators} are also consistent in the sense that for every fixed value $y$ and $t$ with $T=t$ and $Y_T=y$:

\begin{equation}
\begin{aligned}
\lim_{n\rightarrow\infty} \hat{f}_n \left(t\right) = f\left(t\right)\\
\lim_{n\rightarrow\infty} \hat{g}_n \left(y\right) = g\left(y\right);
\end{aligned}
\end{equation}
moreover, $\sqrt{n}\cdot\left( \hat{f}_n\left(t\right)-f\left(t\right)\right)$ and $\sqrt{n}\cdot\left( \hat{g}_n\left(y\right)-g\left(y\right)\right)$ are asymptotically normally distributed with mean zero for every fixed $y$ and $t$.

Hence, FML could help effectively resolve the problems induced by endogeneity. Since the premise of applying FML is the existence of an explicit and parametric expression of the conditional joint probability density function $P$ which can be provided by the Equation \ref{joint density 2nd expression} \footnote{To convert the joint density $P$ in \ref{joint density 2nd expression} to $P_{\alpha}(\,.\,,\,.\,|\beta_{y,0}+\beta_{y}\cdot X,\,\beta_{t,0}+\beta_{t}\cdot X')$, we need to parametrize \ref{joint density 2nd expression} and insert the conditions of explanatory vairabls into it, which will be done in the next sections}, we conclude that RGRST model gives a solid foundation to resolve the endogeneity between charge and LOS.

For the purpose of being illustrative and simple, from now on we will assume that the explanatory variables are introduced into the joint density function $P$ in Equation \ref{joint density 2nd expression} through a linear way, which means the conditional density function $P^{\zeta}_{\alpha}(\,.\,,\,.\,|\beta_{y,0}+\beta_{y}\cdot X,\,\beta_{t,0}+\beta_{t}\cdot X')$ induced by the following random variables: 
\begin{equation}\label{residuals}
\left(\zeta_{Y_T},\, \zeta_{T}\right) := \left(\ln Y_T - \beta_{y,0}-\beta_{y}\cdot X,\,\ln T - \beta_{t,0}-\beta_{t}\cdot X'\right),
\end{equation}
is a joint probability density derived from an RGRST process and invariant for different values of $\beta_{y,0}+\beta_{y}\cdot X$ and $\beta_{t,0}-\beta_{t}\cdot X'$. It turns out that the linear assumption is equivalent to combining the following two conditions together:

(1) The function $f$ and $g$ in Equation \ref{structural equation} has the form of $f(y)=\beta_{f}\cdot \ln T$ and $g(y)=\beta_{g}\cdot \ln Y_T$, meanwhile Equation \ref{structural equation} can be rewritten in the following way: 
\begin{equation}\label{structural equation II}
\begin{aligned}
\ln Y_T =& \beta'_{y,0}+\beta'_{y}\cdot X + \epsilon'_{Y_T}\\
\ln T = & \beta'_{t,0}+\beta'_{t}\cdot X' + \epsilon'_{LOS},
\end{aligned}
\end{equation}
where  
\begin{equation}
\vec{a}'=\begin{pmatrix}1 & -\beta_{f}\\
-\beta_{g} & 1
\end{pmatrix}^{-1}\cdot\,\vec{a}
\end{equation}
holds for $\vec{a}_0  =(\beta_{y},\beta_{t})^T$/$(\beta_{y,0},\beta_{t,0})^T$/$(\epsilon_{Y_T},\beta_{T})^T$ and 
$\vec{a}_0 = (\beta_{y},\beta_{t})^T$/$(\beta_{y,0},\beta_{t,0})^T$/$(\epsilon_{Y_T},\beta_{T})^T$.

(2) The exponential of the residual terms, $\left(e^{\epsilon'_{Y_T}},\,e^{\epsilon'_{T}}\right)$, induces a joint density function derived from an RGRST process as given in Equation \ref{joint density 2nd expression} .

Without loss of generality, we will only consider regression equation \ref{structural equation II} in the following sections\footnote{Notice that it is possible and interesting to assume the joint density $P$ depends on explanatory variables non-linearly. FML can definitely handle those non-linear cases, but introducing too much non-linearity would cause the maximization algorithm less reliable. For the purpose of illustration, only focusing on the linear case should be enough.}.

\subsection{Flexibility of RGRST process}

One big advantage of RGRST process is its flexibility. In fact, we have the following:
\begin{theorem} \label{theorem flexibility}
Fix $\tilde{q}$ with $\tilde{q}\left(0,.\right)\equiv0$, then for
every probability density function $f$ over $[0,\infty)^{2}$, there
always exist a $\tilde{q}_{1}$ function, an initial density function
$p\left(.,0\right)$ and a RGRST process associated with the triple
$\left(p\left(.,0\right),\tilde{q},\tilde{q}_{1}\right)$ as constructed
in Lemma \ref{lemma5} such that the derived joint density of charge and LOS is
given by $f$. Moreover, such a pair of $\left(\tilde{q}_{1},p\left(.,0\right)\right)$
is uniquely determined by the pair of $\left(\tilde{q},f\right)$.
\end{theorem}
The proof of Theorem \ref{theorem flexibility} is given in Appendix \ref{Proof for Theorem flex}.
Theorem \ref{theorem flexibility} implies that RGRST processes provide a very general framework to fit joint distributions over $[0,\infty)^{2}$ in the sense that all probability density function over $[0,\infty)^{2}$ can be achieved as a joint density of charge and LOS derived from some RGRST process. Moreover, the fitting is irrelevant with the choice of $\tilde{q}$. In the other words, no matter what $\tilde{q}$ we choose, it is always possible to fit a given joint density as we want as long as $\tilde{q}_1$ and the initial distribution are chosen properly. As a consequence, we can always choose a simple enough $\tilde{q}$ for the purpose of convenience. Especially, we can choose:
\begin{equation}\label{q_tilde form}
\tilde{q}\left(y,t\right)=a\cdot y,
\end{equation}
with $a>0$.

Using Equation \ref{q_tilde form}, we can express $\tilde{p}$ as below:
\begin{equation}\label{p_tilde}
\tilde{p}\left(y,t\right) = p\left(y\cdot e^{-a\cdot t},0\right)\cdot e^{-a\cdot t}.
\end{equation}
Plugging in \ref{p_tilde} to Equation \ref{joint density 2nd expression} we obtain a way to express the joint density function $P$ completely on the basis of $\tilde{q}_1$ and $p\left(.,0\right)$. Therefore, to parametrize $P$, it is sufficient to parametrize the functions $\tilde{q}_1$ and $p\left(.,0\right)$. This fact facilitates the construction of likelihood function from expression \ref{joint density 2nd expression} and the implementation of FML. 
\section{Full Maximum Likelihood Estimation of RGRST Process}
\subsection{Likelihood Function } \label{likelihood sec}

In this section, we provide a parametric form of the likelihood function that will be estimated in the next section. Formally, the logarithm of the likelihood function is given as below:

\begin{equation} \label{likelihood function}
L:=\sum_{n=1}^{N}\ln\left(\tilde{p}\left(y_{n},t_{n}\right)\cdot\left(-\frac{\partial\tilde{q}_{1}}{\partial y}\cdot\tilde{q}-\frac{\partial\tilde{q}_{1}}{\partial t}\right)\left(y_{n},t_{n}\right)\right)
\end{equation}

By the discussion in the last section, to parametrize $L$, it suffices to choose a specific form for the functions $p\left(.,0\right)$ and $\tilde{q}_{1}$. 

\subsubsection{Principles of Choosing Parametrization of $\tilde{q}_1$ and $p(.,0)$}
Notice that there is not any given procedure to choose a specific parametric form of the functions $\tilde{q}_1$ and $p(.,0)$ neither is there only one unique valid parametric form. For the purpose of being illustrative, we will only choose one \enquote{nice} parametric family to estimate, but our choice does not exclude the existence of other parametric families that are also \enquote{nice} to fit real data. Here, a parametric family of RGRST processes is \enquote{nice} only if it makes the following two principles hold:
\begin{principle}\label{criteria RGRST}
Given Equation \ref{q_tilde form}, the parametric form of $\tilde{q}_1$ and $p(.,0)$ should guarantee Property \ref{Property 1} and \ref{Property 3} hold. 
\end{principle}

\begin{principle}\label{criteria data}
The marginal distributions of charge and LOS derived from the parametric family of RGRST processes should be close to a log-normal distribution and a Phase-Type distribution respectively.
\end{principle}

\begin{principle}\label{criteria mle}

1. For each given profile of parameter values, there is a unique RGRST model corresponding to it. 

2. Different profiles of parameter values should correspond to different RGRST processes and different joint density $P$. 
\end{principle}

Principle \ref{criteria RGRST} just requires the parametric form of $\tilde{q}_1$ and $p(.,0)$ should guarantee the existence of a well-defined RGRST process.
 
Principle \ref{criteria data} requires the marginal distributions derived from the parametric RGRST models should reflect the widely observed fact. That is, the inpatient total charge is approximately following a log-normal distribution for a wide range of health databases and patient groups \cite{gardiner2002longitudinal},\cite{polverejan2003estimating},\cite{Tang2012thesis}, and the hospital LOS can be very well fitted by Coxian Phase-Type distribution \cite{Tang2012thesis}, \cite{faddy2009modeling},\cite{marshall2002modelling},\cite{marshall2007estimating},\cite{marshall2005length}.

The first part of Principle \ref{criteria mle} requires that the underlying RGRST model can be completely identified from the estimated parameter values, which guarantees that we can extract useful information regarding the underlying treatment dynamics that patients experience (stored in the corresponding parametric RGRST model) from charge and LOS data. By the uniqueness part of Theorem \ref{theorem flexibility} and the parametrization we chose for $\tilde{q}$ as in Equation \ref{q_tilde form}, part 1 of Principle \ref{criteria mle} always hold.

Part 2 of Principle \ref{criteria mle} is required by FML method to guarantee the consistency of the estimator, which is verified in Proposition \ref{theorem validation} for the choice of parametric family of the function $\tilde{q}_1$ and $p(.,0)$ as introduced below.

\subsubsection{Parametric Form of $\tilde{q}_1$ and $p(.,0)$}

To avoid discontinuity, we require the support of $p\left(.,0\right)$
to be $[0,\infty)$. In practice, we will choose the probability density function induced by the absolute value of a Cauchy random variable,
i.e.

\begin{equation} \label{initial density}
p\left(y,0\right)=\frac{2}{\pi\gamma\cdot\left(1+\left(\frac{y}{\gamma}\right)^{2}\right)},\,\,\pi,\gamma>0.
\end{equation}

The intuition behind the Equation \ref{initial density} is that most patients only have relatively low charge in the beginning, but it is possible for a small group of outlier patients who are charged with a large amount of money on the first day.

To avoid the subtlety of the requirement in Property \ref{Property 3}
on the relation between $\tilde{q}$ and $\tilde{q}_{1}$, we require that
$\frac{\partial\tilde{q}_{1}}{\partial y}$ and $\frac{\partial\tilde{q}_{1}}{\partial t}$
are non-positive for all points in $\left(0,\infty\right)^{2}$, although they are not for a general RGRST process. With
the non-positive restriction, Property \ref{Property 3} is always satisfied. Moreover, the non-positive requirement reflects such an intuition that the increasing of the total charge and the total length of time that patients stay in hospital are important factors that drive patients leave hospital.
 
In practice, we adopt the following parametric form for $\tilde{q}_{1}$.

\begin{equation} \label{q1_tilde form}
\tilde{q}_{1}\left(y,t\right)=\sum_{n=1}^{N}\theta_{n}\cdot\left(1-PH_{n}\left(t\right)\right)\cdot\left(1-\Phi\left(\frac{\ln\left(y\right)-\mu_{n}}{\sigma_{n}}\right)\right)
\end{equation}
where $\theta_{n}>0$ for each $n$ and satisfies $\sum_{n=1}^{N}\theta_{n}=1$;
$\Phi$ is the standard normal CDF;
$PH_{n}$ is CDF for a Coxian Phase Type
distribution which is of the following general form

\[
PH_{n}\left(t\right)=1-e_{1,d_{n}}\cdot e^{S_{n}\cdot t}\cdot1_{d_{n}}
\]

where $S_{n}$ is a $d_{n}\times d_{n}$ transition matrix characterizing
the Coxian Phase Type process with $d_{n}$ transient stages. $e_{1,d_{n}}$
is a $d_{n}$ dimensional vector with the first entry being $1$ and
all other entries being 0. $1_{d_{n}}$ is the $d_{n}$ dimensional
vector with all entries being $1$. 

The function $\tilde{q}_{1}$ as shown in \ref{q1_tilde form} is just a convex combination
of the product of survival functions of Phase Type distributions and
log-normal distributions. It is easy to check that such defined $\tilde{q}_{1}$ is smooth, with $\frac{\partial\tilde{q}_{1}}{\partial y}$
and $\frac{\partial\tilde{q}_{1}}{\partial t}$ strictly negative
and has range $\left(0,1\right)$. Therefore, the Principle \ref{criteria RGRST} hold for $\tilde{q}_1$ and $p(.,0)$ constructed in Equation \ref{initial density}, \ref{q1_tilde form} and \ref{q_tilde form}.

As mentioned before, the parametric form of $\tilde{q}_1$ and $p(.,0)$ chosen in this paper should be \enquote{nice} in the sense of Principle \ref{criteria RGRST} - \ref{criteria mle}. The following two propositions show that our choice \ref{q1_tilde form} and \ref{initial density} are indeed \enquote{nice}. But notice that the \enquote{niceness} of Equation \ref{q1_tilde construction} and \ref{initial density} does not means there is not any other parametric form for $\tilde{q}_1$ and $p(.,0)$ that is also \enquote{nice} in terms of Principle \ref{criteria RGRST} - \ref{criteria mle}. We chose the Equation \ref{q1_tilde form} and \ref{initial density} partially because they are the first \enquote{nice} example we came up with.

Proposition \ref{theorem validation} verifies that Principle \ref{criteria data} holds for Equation \ref{initial density} and \ref{q1_tilde form}. Its proof is given in Appendix \ref{Proof for Theorem validation}.

\begin{proposition} \label{theorem validation}
Given a $\tilde{q}_{1}$ function of the form in \ref{q1_tilde form}, $\tilde{q}$
of the form in \ref{q_tilde form} and $p\left(.,0\right)$ of the form in \ref{initial density}, the
joint density function of charge and LOS can be expressed as:

\begin{equation}
\label{joint density for estimation}
\begin{aligned}
p_{Y_{T},T}\left(y,t\right)=&\frac{2}{\pi\gamma\cdot\left(1+\left(\frac{y\cdot\exp\left(-a\cdot t\right)}{\gamma}\right)^{2}\right)}\times\\
&
\begin{split} 
 \sum_{n=1}^{N}\theta_{n}\cdot\left(e_{1,d_{n}}\cdot e^{\left(S_{n}-a\right)\cdot t}\cdot1_{d_{n}}\cdot\frac{\exp\left(-\frac{\left(\ln\left(y\right)-\mu_{n}\right)^{2}}{2\sigma_{n}}\right)}{\sqrt{2\pi}\sigma_{n}}\cdot a \right.\\
\left. - e_{1,d_{n}}\cdot e^{\left(S_{n}-a\right)\cdot t}\cdot S_{n}\cdot1_{d_{n}}\cdot\left(1-\Phi\left(\frac{\ln\left(y\right)-\mu_{n}}{\sigma_{n}}\right)\right)\right)
\end{split}
\end{aligned}
\end{equation}

Moreover, the marginal distribution for LOS is close to a Coxian
Phase Type distribution with difference controlled by $C\cdot\exp\left(-a\cdot t\right)$
(constant $C$ depends on the Phase Type components $PH_{i}$'s and
log-normal components $\left(\mu_{i},\sigma_{i}\right)$'s ); the
marginal density function of total charge has its right tail asymptotically
equivalent to a log-normal density function as $y\rightarrow\infty$.\end{proposition}

In the next proposition, we shall verify the part 2 of Principle \ref{criteria mle}. Here we consider RGRST models parametrized through Equation \ref{q_tilde form}, \ref{q1_tilde form} and \ref{initial density} together with the regression equation specified in \ref{structural equation II}. Notice that when regression equation \ref{structural equation II} is involved, the parameter $\gamma$ in \ref{initial density} and $a$ in \ref{q_tilde form} become redundant: we can set them to be the constant, 1.

\begin{proposition} \label{identification }
For each patient group identified by a
fixed profile of explanatory variables $\left(x,x'\right)$ in the regression equation \ref{structural equation II}, the underlying RGRST
process of this patient group can be uniquely identified under the
assumption that the functions $p_{x,x'}\left(.,0\right)$, $\tilde{q}_{x,x'}$ and
$\tilde{q}_{1,x,x'}$ associated with the unique RGRST
process have the form given in equations \ref{initial density}, \ref{q_tilde form} and \ref{q1_tilde form} respectively.
Moreover, $p_{x,x'}\left(.,0\right)$, $\tilde{q}_{x,x'}$ and $\tilde{q}_{1,x,x'}$
can be expressed as follows:

\begin{eqnarray*}
p_{x,x'}\left(y,0\right) & = & \frac{2}{\left(1+\left(\frac{y}{\exp\left(\beta_{0,c}+\beta_{c}\cdot x\right)}\right)^{2}\right)\cdot\pi\cdot\exp\left(-\beta_{0,c}-\beta_{c}\cdot x\right)}\\
\tilde{q}_{x,x'}\left(y,t\right) & = & \frac{y}{\exp\left(\beta_{0,los}+\beta_{los}\cdot x'\right)}\\
\tilde{q}{}_{1,x,x'}\left(y,t\right) & = & \tilde{q}_{1}\left(\frac{y}{\exp\left(\beta_{0,y}+\beta_{y}\cdot x\right)},\frac{t}{\exp\left(\beta_{0,los}+\beta_{los}\cdot x'\right)}\right)
\end{eqnarray*}

where $\tilde{q}_{1}$ is given as in Equation \ref{q1_tilde form}.\end{proposition}

Proof for Proposition \ref{identification } is in Appendix \ref{Proof for Proposition id}. Proposition \ref{identification } shows that the correspondence between the parameter space and the parametric family of functions $\tilde{q}$, $\tilde{q}_1$ and $p(.,0)$ is one-to-one, which verifies the part 2 of Principle \ref{criteria mle}.

\subsubsection{Expression for Log-Likelihood Function}
Combining the regression equation \ref{structural equation II}, the joint density function
given in \ref{joint density for estimation} and the log-likelihood function in \ref{likelihood function}, we obtain the parametric form of the
log-likelihood function as below:

\begin{equation}\label{likelihood for estimation}
\Scale[0.8]{
\begin{aligned}
L & :=  \sum_{k=1}^{K}\ln\left(\frac{2\cdot\exp\left(-\beta_{0,los}-\beta_{los}\cdot x_{k}'\right)\cdot\exp\left(-\beta_{0,c}-\beta_{c}\cdot x_{k}\right)}{\pi\left(1+\left(\frac{y_{k}\cdot\exp\left(-t_{k}/\exp\left(\beta_{0,los}+\beta_{los}\cdot x_{k}'\right)\right)}{\exp\left(\beta_{0,c}+\beta_{c}\cdot x_{k}\right)}\right)^{2}\right)}\right)+\\
&
\begin{split} 
\sum_{k=1}^{K}\ln\left(
\begin{split}
\sum_{n=1}^{N}\theta_{n}\cdot\left(e_{1,d_{n}}\cdot e^{\frac{\left(S_{n}-1\right)}{\exp\left(\beta_{0,los}+\beta_{los}\cdot x_{k}'\right)}\cdot t_{k}}\cdot1_{d_{n}}\cdot\frac{\exp\left(-\frac{\left(\ln\left(y_{k}\right)-\beta_{0,c}-\beta_{c}\cdot x_{k}-\mu_{n}\right)^{2}}{2\sigma_{n}}\right)}{\sqrt{2\pi}\sigma_{n}} -\right.\\
\left.e_{1,d_{n}}\cdot e^{\frac{\left(S_{n}-1\right)}{\exp\left(\beta_{0,los}+\beta_{los}\cdot x_{k}'\right)}\cdot t_{k}}\cdot S_{n}\cdot1_{d_{n}}\cdot\left(1-\Phi\left(\frac{\ln\left(y_{k}\right)-\beta_{0,c}-\beta_{c}\cdot x_{k}-\mu_{n}}{\sigma_{n}}\right)\right)\right)
\end{split}
\right)
\end{split}
\end{aligned}}
\end{equation}
and the following constrained maximization problem that we have to
solve in order to get the estimated parameters:

\begin{equation}
\label{opt problem}
\begin{aligned}
\max & L\left(\theta;\beta;LN;s\right)\\
s.t.\, & \sum_{k=1}^{K}\theta_{k}=1\\
 & \theta_{k}\geq0 & k=1,\dots K\\
 & \sigma_{k}\geq0 & k=1,\dots,K
\end{aligned}
\end{equation}
where the four families of parameters to be solved are the weight vector
$\theta=\left(\theta_{1},\dots,\theta_{K}\right)$, the regression
coefficients $\Scale[0.8]{\beta=\left(\beta_{0,c},\dots,\beta_{\left|x\right|,c},\beta_{0,los},\dots,\beta_{\left|x'\right|,los}\right)}$
and parameter vectors for the log-normal components $\Scale[0.7]{LN=\left(\mu_{1},\dots,\mu_{K},\sigma_{1},\dots,\sigma_{K}\right)}$,
and the parameter vectors for the Phase Type components $s=\left(s_{1},\dots,s_{K}\right)$
where each $s_{k}$ is a $2d_{k}-1$ dimensional vector which determines
the transition matrix $S_{k}$ in the following way:

\[
S_{k,ij}=\begin{cases}
-\exp\left(s_{k,2i-1}\right) & i=j\\
\exp\left(s_{k,2i}\right) & j=i+1\\
0 & j\not=i\, or\, i+1
\end{cases}
\]

Here, we adopt the transformation used in \cite{esparza2010maximum}
to remove the boundary requirement for entries of each $S_{k}$.

We solve the maximization problem \ref{opt problem} using Python-Scipy optimization package.
The details of the data and estimation results are reported in section 3.2.1. 

\subsubsection{Model Dimension}

Notice that the dimension of parametric RGRST models is not determined by now, and it is affected by the parameter $N$ in Equation \ref{q1_tilde form} that represents the number of components appearing in the convex combination of the parametric form of $\tilde{q}_1$. For every fixed $N$, the model dimension is also affected by the dimension, $d_i$, of transition matrix of the $i$th Phase-Type component with $i \in \left\{1,\dots,N\right\}$.

The value of the dimension parameters $\left(N,d_1,\dots,d_N\right)$ characterizes the complexity of the joint density function $P$. In fact, from Equation \ref{joint density for estimation},  we can conclude that the dimension parameter $N$ measures the complexity of the slice of the density function $P\left(.,t\right)$ (viewed as a function on variable $y$) for each fixed time $t$. Roughly speaking, if we measure the complexity of a density function through the number of modalities that it has, then using the uni-modal property of the log-normal density function we can verify that the value of $N$ gives an upper bound of the number of modalities that the family of functions $\left\{P\left(.,t\right):\,t\in \left[0,\infty\right)\right\}$ could have.
In contrast, the value of $\left(d_1,\dots,d_N\right)$ describes the complexity of the function $P\left(y,.\right)$ on variable $t$ for each fixed charge level $y$. But unlike the log-noraml distributions, the Coxian Phase-Type distribution is not uni-modal in general, there is not a simple way to describe how the values of each $d_i$ could affect the complexity of $\left\{P\left(y,.\right):\,t\in\left[0,\infty\right)\right\}$.

It is necessary to specify the dimension parameters $\left(N,d_1,\dots,d_N\right)$ in order to completely specify a parametric RGRST model. But due to their discrete essence, we have to deal with them in a different way from the continuous parameters in estimation. 

There are a couple of different ways to select the model dimension. Reversible Jump Markov Chain Monte Carlo method (RJMCMC) \cite{green1995reversible} provides a framework to automate the estimation of the discrete dimensional parameters together with continuous parameters. Xiaoqing \cite{Tang2012thesis} applied RJMCMC and Coxian Phase-Type distributions to fit LOS data, which enabled her to fit the transition matrix and estimate the most possible dimension of a Coxian Phase-Type model simultaneously. 

Alternatively, as proposed in \cite{schwarz1978estimating}, one can estimate a class of models (with different dimensions) by FML and compute the BIC (or AIC) from the likelihood function. The final model dimension is determined through comparing the scores of BIC (or AIC) of models with different dimension. Faddy \cite{faddy2009modeling} applied this method to determine the dimension of a Phase-Type model that could fit LOS data best.\\

\begin{minipage}{1\textwidth}
\begin{center}
\captionof{figure}{Joint Histogram of Charge and LOS SPARCS 2013 (Entire Database)}\label{fig:joint density whole}
\includegraphics[width=10cm,height=5cm]{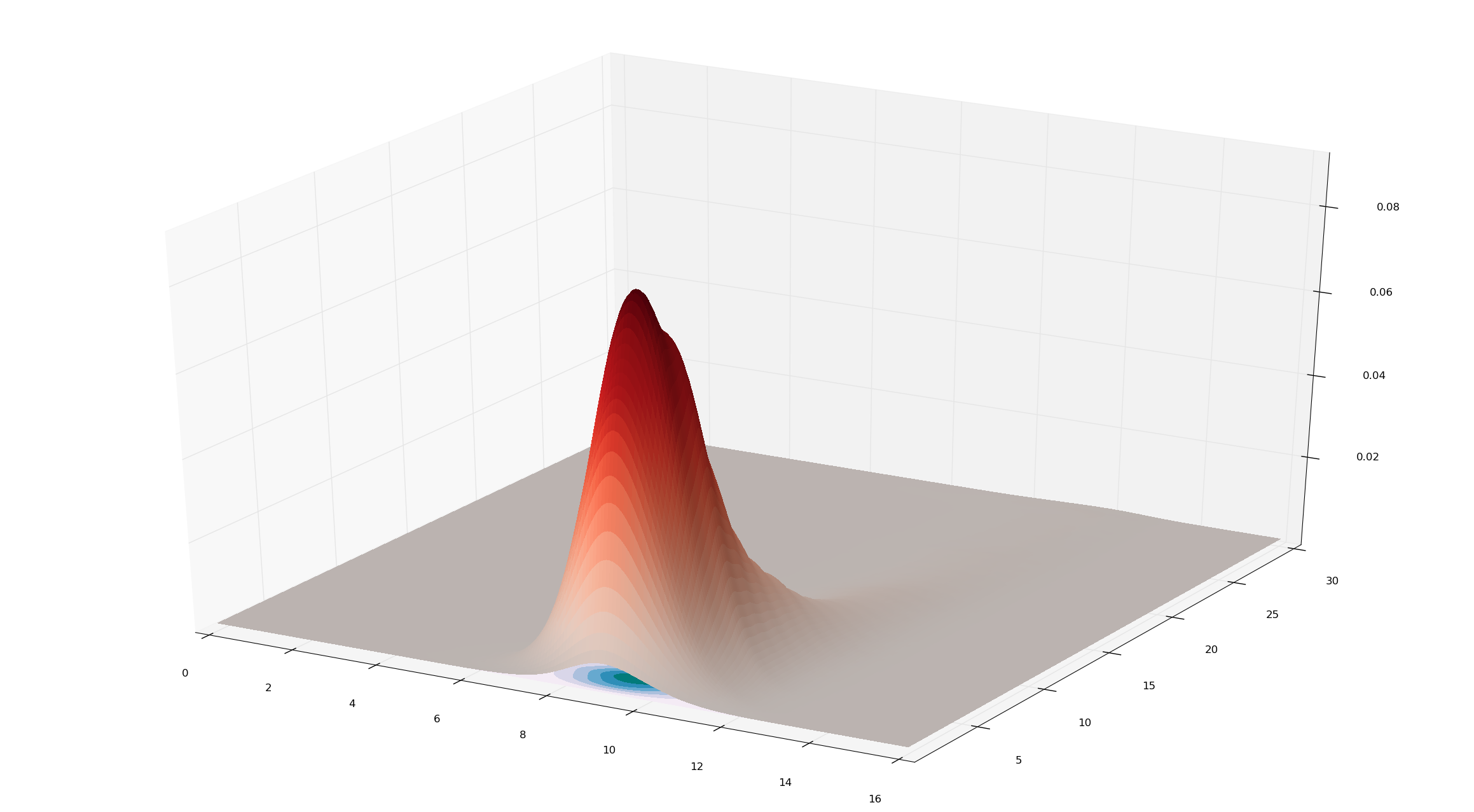}
\end{center}
\end{minipage}\\

In this paper, we only consider those parametric RGRST models with $N = 1$, because through plotting the histogram of charge and LOS (as shown in Figure \ref{fig:joint density whole}), we found that the modality structure of the family of functions $\left\{P_{e}\left(.,t\right):\,t\in\left[0,\infty\right)\right\}$ ($P_{e}$ is the empirical density function given by the histogram in Figure \ref{fig:joint density whole}) is consistent with the parametric family of RGRST models with $N=1$. For the dimension parameter $d_1$, we will only consider the case $d_1 \leq 5$ for simplicity.

\subsection{Estimation}

\subsubsection{Patient Sample}

In this section, we apply the RGRST process to fit real inpatient data
and perform the regression analysis. We use inpatient
data from New York State's Statewide Planning and Research
Cooperative System 2013 (SPARCS 2013).

SPARCS is a system initially created to collect information on discharges
from hospitals within New York State. SPARCS currently collects patient
level detail on patient characteristics, diagnoses and treatments,
services, and charges for each hospital inpatient stay and outpatient visit; and each ambulatory surgery and outpatient services visit to
a hospital extension clinic and diagnostic and treatment center licensed to provide ambulatory surgery services. In 2013, the SPARCS contains
nearly 2.5 million inpatient discharges from 218 facilities and 58
counties in New York State. Patient demographics in the SPARCS include
age group at admission, gender,
race, source of payment and zip code. Patient clinical characteristics
include type of admission, diagnosis codes (MDC code, DRG code, CCS
diagnosis code etc.) and treatment procedures undergone (CCS Procedure
Code). In this article, our aim is at illustrating the methodology proposed in previous sections, so we intend to choose explanatory variables as simple as possible. Moreover, we are most interested in the effect of diagnosis-related covariates on charge and LOS, therefore the covariates we choose for estimation are MDC code, Severity of Illness Code and Risk of Mortality.

Intuitively, patients who have more severe illness condition and higher risk of mortality tend to stay in hospital for longer time and get charged more money.
SPARCS 2013 verified this intuition. Patients with the most severe conditions (APR Severity
of Illness Code = 4) have an extremely long mean LOS (almost 17 days)
and the highest mean charge (\$210806). Similarly to the severity of illness,
patients with highest risk to die has the longest mean LOS (almost
15 days) and highest mean charge (\$200746).  

Descriptive statistics of SPARCS 2013 with respect to the chosen covariates are presented in Table \ref{table: 3.1}.

\begin{minipage}{\linewidth}
\begin{center} 
\captionof{table}{Descriptive statistics of SPARCS 2013 (SET 1)}\label{table: 3.1}
\centering
\resizebox{\columnwidth}{!}{%
\begin{tabular}{cccccccc}
\hline 
\hline 
Characteristics & Group & N(\%) & Sample\_N(\%) & LOS(SD) & Sample\_LOS(SD) & Charge(SD) & Sample\_Charge(SD)\tabularnewline
\hline 
All Patients &  & 2418874(100) & 5000(100) & 5.46(8.11) & 5.51(8.16) & 36931.77(68973.47) & 36861.8(67053.64)\tabularnewline
MDC & 0.0 & 17.0(0.0) &  & 11.0(24.69) &  & 102910.82(280754.64) & \tabularnewline
 & 1.0 & 142651.0(5.9) & 298.0(5.96) & 5.7(8.69) & 5.01(6.14) & 46962.08(83724.59) & 41911.53(50501.65)\tabularnewline
 & 2.0 & 4138.0(0.17) & 13.0(0.26) & 3.62(5.18) & 3.38(1.89) & 27185.04(37576.18) & 28478.72(22611.85)\tabularnewline
 & 3.0 & 32743.0(1.35) & 72.0(1.44) & 3.59(5.59) & 2.81(2.72) & 29468.92(50592.3) & 22093.15(20516.67)\tabularnewline
 & 4.0 & 206374.0(8.53) & 425.0(8.5) & 5.81(7.64) & 5.42(6.99) & 37165.6(64478.26) & 35254.34(48416.02)\tabularnewline
 & 5.0 & 320765.0(13.26) & 655.0(13.1) & 4.78(6.58) & 4.68(5.4) & 50065.14(84839.8) & 48514.49(67896.89)\tabularnewline
 & 6.0 & 211325.0(8.74) & 461.0(9.22) & 5.11(6.63) & 5.56(7.16) & 35785.32(54820.39) & 37176.32(45615.65)\tabularnewline
 & 7.0 & 65928.0(2.73) & 116.0(2.32) & 5.6(6.96) & 4.91(4.31) & 42718.49(78816.64) & 34176.78(38341.19)\tabularnewline
 & 8.0 & 201134.0(8.32) & 419.0(8.38) & 4.91(5.95) & 5.0(5.35) & 50655.45(55819.15) & 50609.01(45334.44)\tabularnewline
 & 9.0 & 66120.0(2.73) & 136.0(2.72) & 4.6(5.95) & 5.07(8.42) & 28073.74(37308.49) & 28829.12(29869.57)\tabularnewline
 & 10.0 & 74993.0(3.1) & 171.0(3.42) & 3.97(5.83) & 4.05(4.72) & 28568.47(43837.41) & 27236.6(30456.03)\tabularnewline
 & 11.0 & 103597.0(4.28) & 221.0(4.42) & 5.43(6.75) & 5.09(5.1) & 36812.91(53368.81) & 33884.47(38131.54)\tabularnewline
 & 12.0 & 11181.0(0.46) & 21.0(0.42) & 3.44(6.27) & 4.81(10.56) & 30593.31(30945.72) & 39233.29(46533.28)\tabularnewline
 & 13.0 & 31682.0(1.31) & 57.0(1.14) & 3.13(5.23) & 2.47(2.03) & 28998.31(33592.18) & 31389.63(20325.52)\tabularnewline
 & 14.0 & 257203.0(10.63) & 504.0(10.08) & 2.91(2.54) & 2.88(2.47) & 16435.92(17226.17) & 16714.7(18104.7)\tabularnewline
 & 15.0 & 236599.0(9.78) & 439.0(8.78) & 3.78(7.99) & 4.06(7.8) & 17912.83(85865.5) & 18682.72(72830.49)\tabularnewline
 & 16.0 & 37899.0(1.57) & 92.0(1.84) & 5.01(6.87) & 4.77(3.79) & 37100.38(83604.25) & 36537.56(52336.47)\tabularnewline
 & 17.0 & 22289.0(0.92) & 55.0(1.1) & 9.57(12.73) & 9.38(11.59) & 87130.44(139632.35) & 81519.0(128268.96)\tabularnewline
 & 18.0 & 108416.0(4.48) & 224.0(4.48) & 9.09(10.7) & 10.24(15.07) & 63423.77(99592.33) & 80804.11(200106.77)\tabularnewline
 & 19.0 & 116683.0(4.82) & 245.0(4.9) & 12.94(16.11) & 12.62(17.7) & 34162.28(57058.45) & 32507.77(49653.33)\tabularnewline
 & 20.0 & 75432.0(3.12) & 170.0(3.4) & 6.34(7.45) & 6.6(7.65) & 17400.15(23797.61) & 17228.44(20575.0)\tabularnewline
 & 21.0 & 30203.0(1.25) & 71.0(1.42) & 4.29(7.18) & 4.77(10.44) & 31248.52(64435.64) & 33845.59(75320.46)\tabularnewline
 & 22.0 & 1929.0(0.08) & 2.0(0.04) & 9.06(13.5) & 8.0(2.83) & 79337.2(184652.6) & 51080.31(29187.33)\tabularnewline
 & 23.0 & 46924.0(1.94) & 106.0(2.12) & 10.87(10.27) & 11.18(8.88) & 46721.27(52350.27) & 45356.56(37999.3)\tabularnewline
 & 24.0 & 8733.0(0.36) & 20.0(0.4) & 8.6(11.36) & 8.55(8.81) & 57383.57(105543.15) & 40839.06(39649.57)\tabularnewline
 & 25.0 & 3916.0(0.16) & 7.0(0.14) & 10.77(12.01) & 11.0(7.44) & 103841.73(118285.21) & 73790.14(54114.8)\tabularnewline
Severity & 0.0 & 40.0(0.0) &  & 6.35(16.4) &  & 47710.78(186214.68) & \tabularnewline
 & 1.0 & 881300.0(36.43) & 1760.0(35.2) & 3.09(3.97) & 3.02(3.37) & 20164.74(25917.49) & 20249.04(23995.3)\tabularnewline
 & 2.0 & 929347.0(38.42) & 1939.0(38.78) & 4.96(6.89) & 5.16(7.76) & 30512.25(37884.57) & 30602.25(38507.07)\tabularnewline
 & 3.0 & 479712.0(19.83) & 1048.0(20.96) & 7.73(8.46) & 7.57(7.68) & 51935.05(65352.31) & 51307.28(61645.27)\tabularnewline
 & 4.0 & 128475.0(5.31) & 253.0(5.06) & 16.83(18.2) & 17.06(18.36) & 142361.38(210806.88) & 140564.83(210012.56)\tabularnewline
Mortality & Extreme & 106154.0(4.39) & 210(4.2) & 14.96(16.66) & 13.81(15.11) & 129939.83(200746.65) & 114408.34(172257.11)\tabularnewline
 & Major & 311482.0(12.88) & 692(13.84) & 8.69(10.14) & 8.51(9.96) & 61247.22(92604.92) & 64073.56(108815.34)\tabularnewline
 & Minor & 1482115.0(61.27) & 3007(60.14) & 4.03(6.16) & 4.02(6.31) & 24133.09(33016.21) & 23905.06(31377.21)\tabularnewline
 & Moderate & 519083.0(21.46) & 1091(21.82) & 5.67(7.03) & 6.12(7.96) & 39863.27(55375.06) & 40386.64(51083.0)\tabularnewline
\hline 
\hline 
 &  &  &  &  &  &  & \tabularnewline
\end{tabular}%
}
\end{center}
\end{minipage}

\subsubsection{Estimation of RGRST Regression Model}

Due to the huge data size and limited computation power, our estimation
is based on a subset with size 5000 of SPARCS 2013 obtained through uniform random
sampling. Descriptive statistics of the samples are summarized
in Tables \ref{table: 3.1}, (Sample\_ is used to indicate the statistics
computed for samples). Through direct comparison of statistics computed
for samples and for the entire database, we find that our samples
copy the statistical properties of the entire database quite well. 

After preliminary goodness-of-fit analyses, the RGRST model we finally choose for regression has only one component appearing in the convext combination in Equation \ref{q1_tilde form} and the corresponding dimension of Phase-Type transition matrix is 1 as well. That means only three parameters are needed to characterize the underlying RGRST process, two for the log-normal component ($\mu$ and $\sigma$) and one for the Phase-Type component ($s$). Later, we will call this model RGRST (1:1) with the first \enquote{1} representing the number of summand in Equation \ref{q1_tilde form} and the second \enquote{1} denoting the dimension of the corresponding Phase-Type component. The fitting result for both of the marginal and joint densities of log-charge ($\zeta_{Y_T}$) and LOS ($e^{\zeta_T}$)\footnote{$\exp\left(
\zeta_{Y_T}\right)$ and $\exp\left(\zeta_T\right)$ are given as in \ref{residuals} which are just a re-scale of charge and LOS with the scale determined by the value of explanatory vairables through regression equation \ref{structural equation II}.} are plotted in Figure \ref{fig:marginal} and \ref{fig:joint}, where the estimated value of $(\mu, \sigma, s)$ equals to $(2.96, 0.81, 0.02)$. The fitting statistics we consider here is the Pearson's Chi-square, the Chi-square statistics and the associated Pvalues are (0.032,1.0) and (0.365,1.0) for the marginal distributions of $\zeta_{Y_T}$ and $e^{\zeta_T}$ respectively. The computation of Pearson's Chi-square is based on a 200-fold partition of the range of $\zeta_{Y_T}$ (:= $[-10,10]$) and a 30-fold partition of the range of $e^{\zeta_T}$ (:= $[0,30]$). For the joint distribution, the Chi-sqaure statistics and Pvalue is (0.43,1.0), where a $(200\ast30)$-fold product partition is used to compute the joint histogram. From both of the Chi-square statistics and the fitting plot, it is easy to see that the RGRST (1:1) could generate very good fitting to both of the marginal and joint distributions of charge and LOS. The good fitting also validates the properness of using the joint density function derived from RGRST (1:1) for the FML estimation of Equation \ref{structural equation II}.

\begin{minipage}\linewidth
\begin{center}
\captionof{figure}{Marginal Fitting of Log-charge and LOS by RGRST Model}\label{fig:marginal}
\includegraphics[width=9cm,height=4.5cm]{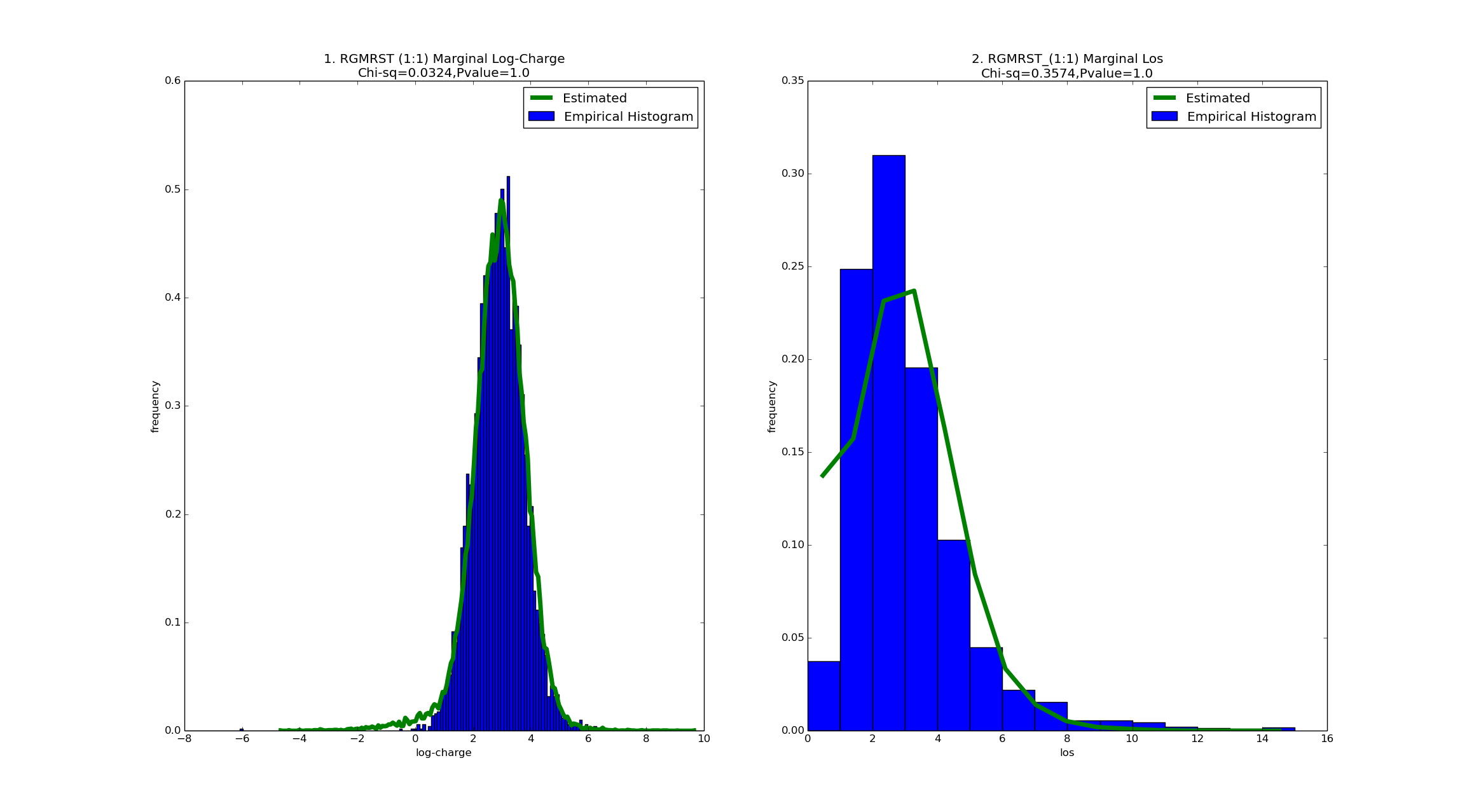}
\includegraphics[width=9cm,height=4.5cm]{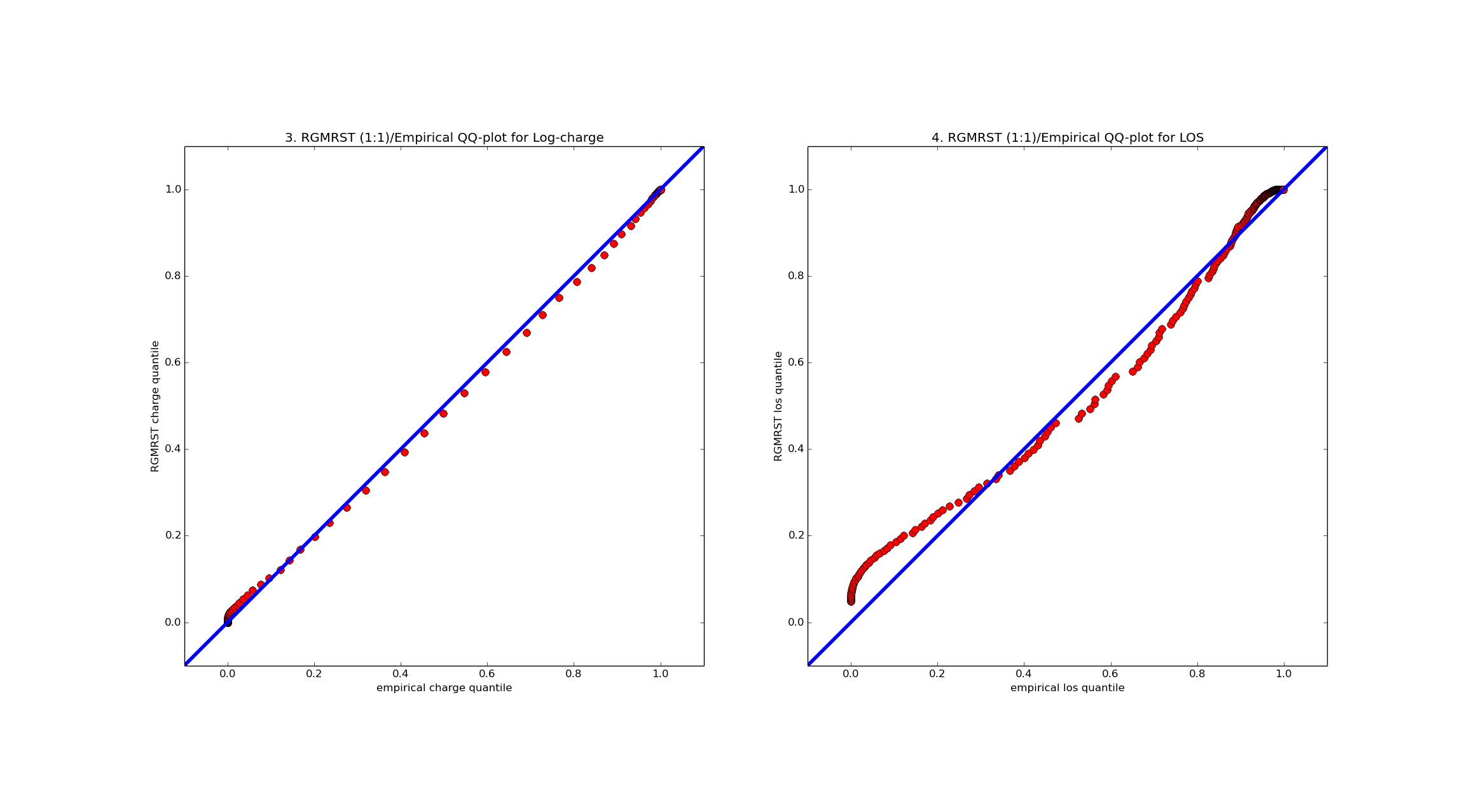}
\includegraphics[width=9cm,height=4.5cm]{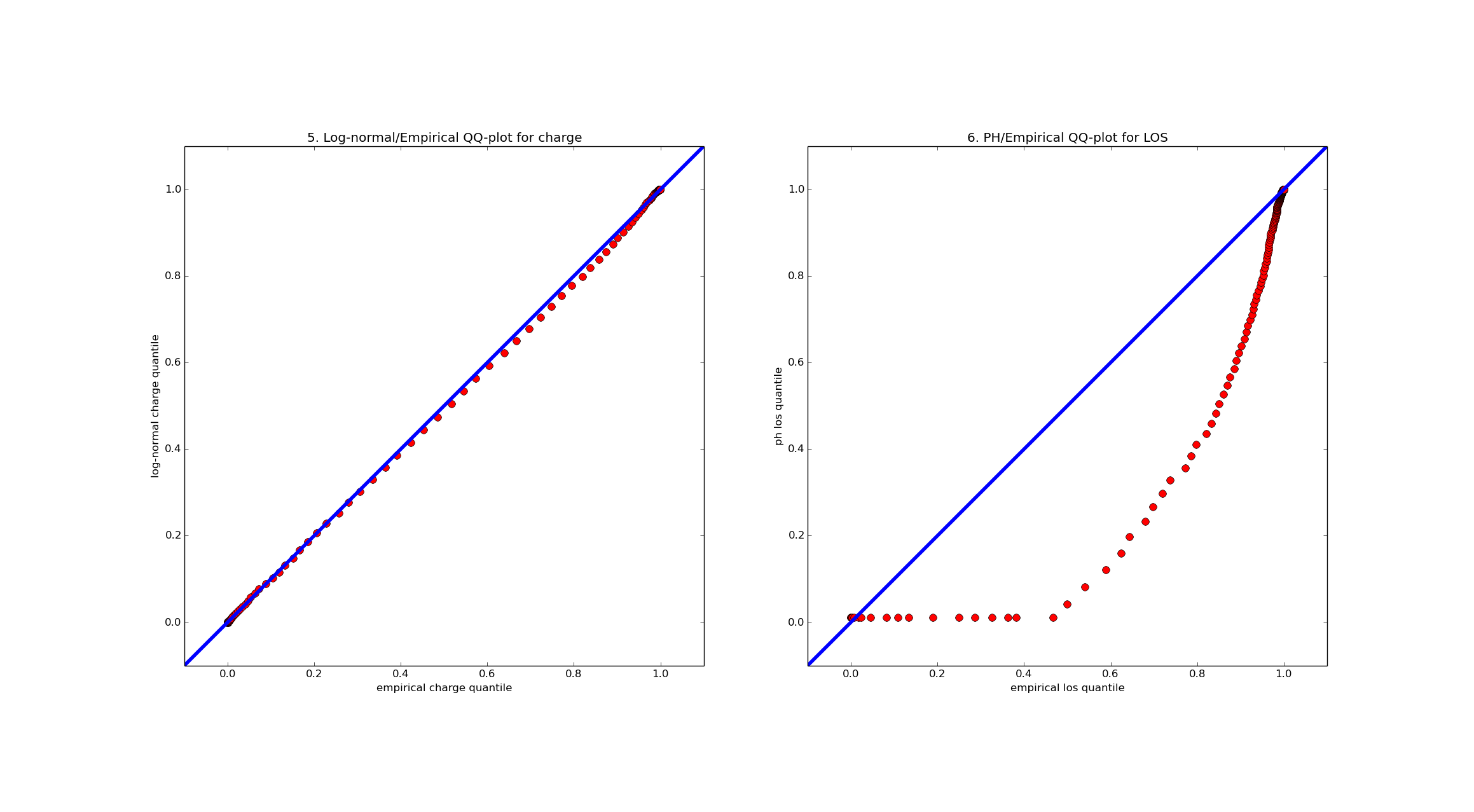}
\end{center}
\end{minipage}
\begin{minipage}\linewidth
\begin{center}
\captionof{figure}{Joint Fitting to Log-charge and LOS by RGRST (1:1)}\label{fig:joint}
\includegraphics[width=14cm,height=7cm]{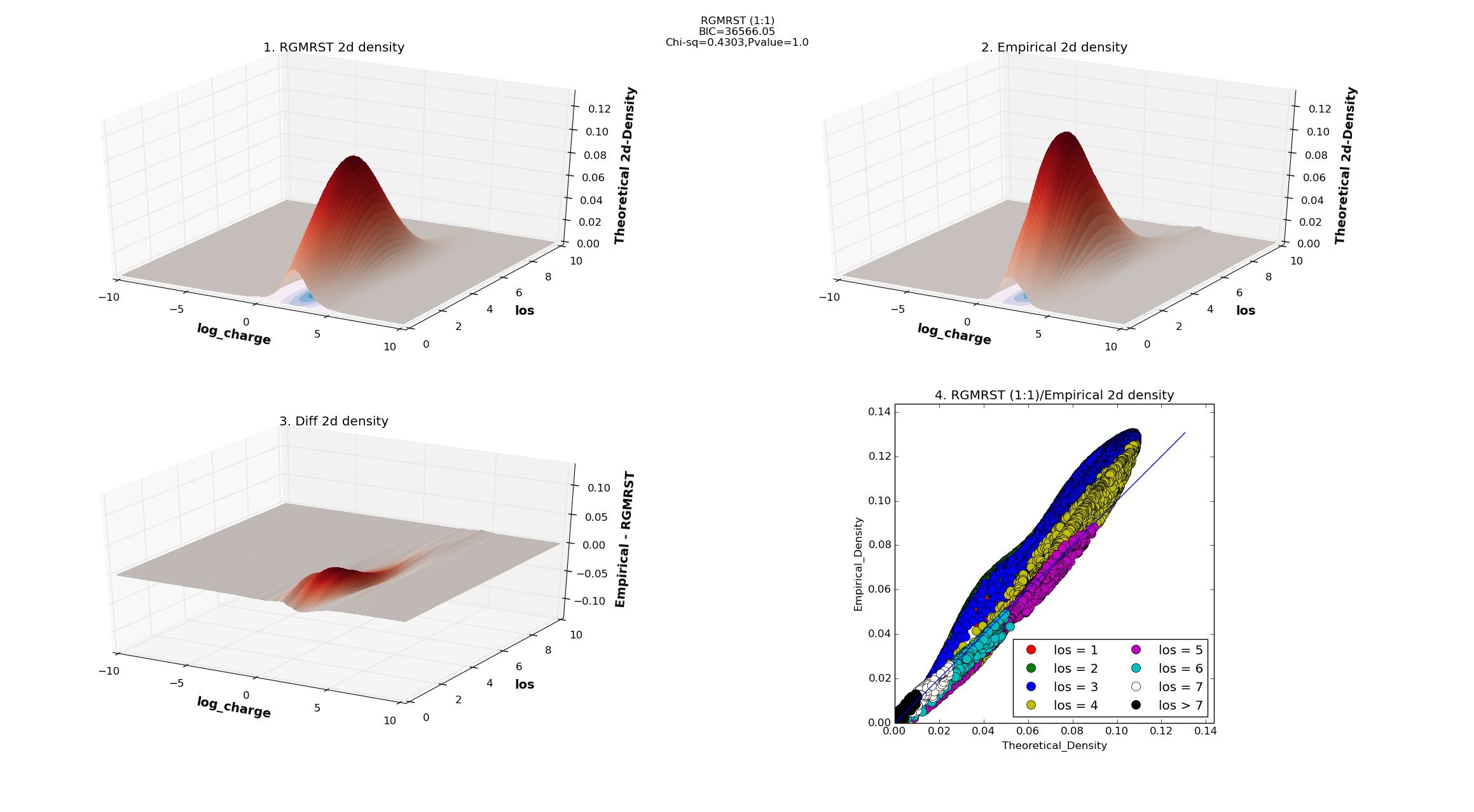}
\begin{minipage}{1\textwidth}
{\footnotesize Plot 1 is the joint density of log-charge and LOS derived from RGRST (1:1). Plot 2 is the empirical joint density obtained from Gaussian kernel density estimation (KDE) with kernel width 0.15 for log-charge and 1 for LOS. Plot 3 is obtained from subtracting Plot 1 from Plot 2. Plot 4 is the KDE density versus the RGRST (1:1) density evaluated at all 5000 samples.\par}
\end{minipage}
\end{center}
\end{minipage}

For comparison, we plot a log-normal fitting to the marginal charge distribution and a Coxian Phase-Type fitting (\cite{Tang2012thesis}) to the marginal LOS distribution in Figure \ref{fig:marginal} (plot 5 and 6). The Phase-Type distribution used here has a 4-dimensional transition matrix, which generates the besting fitting in preliminary goodness-of-fit analyses according to the BIC scores, which is consistent with Xiaoqing's result in \citep{Tang2012thesis}.

To avoid \enquote{overfitting}, we perform an out-sample fitting based on the in-sample estimated parameters. Here the out-sample is chosen to be the complement dataset to the 5000 in-sample within SPARCS 2013, which contains millions of records. The Pearson's Chi-sqaures are computed based on the same partitions as for the in-sample case. The statistics and Pvalues for the out-sample are (0.196,1.0) for $\zeta_{Y_T}$, (0.320,1.0) for $e^{\zeta_T}$ and (0.333,1.0) for the joint. Through comparing the out-sample and in-sample Chi-squares, we see that the estimated Chi-squares in these two cases are not significantly different, which implies the robustness of our fitting. The out-sample fitting plots are presented in Appendix 4.

Comparing the plot 4 and 6 in Figure \ref{fig:marginal}, it is easy to see that the marginal LOS fitting of a 4-phase Coxian Phase-Type distribution (with seven parameters used) is not better than the fitting generated by a RGRST process with only one Phase-Type and log-normal component (with only three parameters used). Moreover, only based on Phase-Type distribution, we cannot generate fitting to charge, nor fitting to the joint of charge and LOS. Therefore, we believe that the RGRST model can provide a more efficient method to fit the empirical charge and LOS data than its competitor, Phase-Type model.

Corresponding to RGRST (1:1), the estimated values of regression
coefficients and the associated P-values are reported in Table
\ref{table: SET1 reg coef sparcs}. Due to the asymptotic normality of maximum likelihood
estimators, P-values in Table \ref{table: SET1 reg coef sparcs} are computed from the
corresponding Fisher information (\cite{amemiya1985advanced}) of the log-likelihood function \ref{likelihood for estimation}.

From Table \ref{table: SET1 reg coef sparcs}, the sign and scale of estimated regression coefficients
coincide with our intuition and the descriptive statistics of SPARCS 2013.

\begin{minipage}{\linewidth}
\begin{center}
\captionof{table}{Estimated Regression Coefficients for RGRST (1:1)}\label{table: SET1 reg coef sparcs}
\resizebox{6cm}{3.25cm}{%
\begin{tabular}{c|ll}
\hline 
\hline
Regressors & Log-Charge(P-values) & LOS(P-values) \tabularnewline
\hline
Intercept & 6.33(0.0074) & -0.1127($<$0.0001) \tabularnewline
MDC\_1 & -0.0375($<$0.0001) & -0.183($<$0.0001) \tabularnewline
MDC\_2 & 0.2094($<$0.0001) & 0.4476($<$0.0001) \tabularnewline
MDC\_3 & -0.2843($<$0.0001) & -0.4269($<$0.0001) \tabularnewline
MDC\_4 & -0.3782($<$0.0001) & -0.2716($<$0.0001) \tabularnewline
MDC\_5 & -0.0179($<$0.0001) & -0.2933(0.0143) \tabularnewline
MDC\_6 & -0.1355($<$0.0001) & -0.175($<$0.0001)\tabularnewline
MDC\_7 & -0.0365($<$0.0001) & -0.1019($<$0.0001) \tabularnewline
MDC\_8 & 0.4908(0.5811) & 0.0098($<$0.0001) \tabularnewline
MDC\_9 & -0.2766($<$0.0001) & -0.211($<$0.0001) \tabularnewline
MDC\_10 & -0.2964($<$0.0001) & -0.3685($<$0.0001) \tabularnewline
MDC\_11 & -0.4415($<$0.0001) & -0.3699($<$0.0001) \tabularnewline
MDC\_12 & 0.0066($<$0.0001) & -0.4127($<$0.0001) \tabularnewline
MDC\_13 & 0.0036($<$0.0001) & -0.3437(0.0836) \tabularnewline
MDC\_14 & -0.5361($<$0.0001) & -0.2767($<$0.0001) \tabularnewline
MDC\_15 & -1.1669($<$0.0001) & -0.1351($<$0.0001) \tabularnewline
MDC\_16 & -0.1704($<$0.0001) & -0.1607($<$0.0001) \tabularnewline
MDC\_17 & 0.2109($<$0.0001) & 0.0659($<$0.0001) \tabularnewline
MDC\_18 & -0.1686($<$0.0001) & -0.1197($<$0.0001)\tabularnewline
MDC\_19 & 0.0887($<$0.0001) & 1.0046($<$0.0001) \tabularnewline
MDC\_20 & -0.5517($<$0.0001) & 0.1839($<$0.0001)\tabularnewline
MDC\_21 & -0.3343($<$0.0001) & -0.2228($<$0.0001) \tabularnewline
MDC\_22 & 0.1158($<$0.0001) & 0.0909($<$0.0001) \tabularnewline
MDC\_23 & 0.0301($<$0.0001) & 0.5076($<$0.0001) \tabularnewline
MDC\_24 & -0.3241($<$0.0001) & -0.1595($<$0.0001) \tabularnewline
Mortality & 0.1529(0.0264) & 0.2036($<$0.0001)\tabularnewline
Severity & 0.423(0.3314) & 0.3469($<$0.0001) \tabularnewline
\hline 
\hline
\end{tabular}%
}
\end{center}
\end{minipage}

\subsubsection{Implications to Patient's Behavioral Patterns}

One important by-product of RGRST processes is that from it we can get useful inference of the behavioral patterns that patient and/or doctor displays when they make discharge decision in reaction to the change of total charge and the total length of time that patient has spent in hospital (which is characterized by the conditional probability $\tilde{q}_1$, whose parametric expression can be estimated from FML). Moreover, we can get a detailed characterization of how the actual charge is accumulated over time through the time-dependent density function $p\left(y,t\right):=
Prob\left(Y_t\in\left[y,y+dy\right)\right)$
whose expression is given in Equation \ref{time dependent density 2} and its parametric form can be derived from the FML estimators in the previous section as well.

\begin{minipage}\linewidth
\begin{center}
\captionof{figure}{FML Estimation of Function $p$}
\begin{subfigure}[b]{5cm}            
\frame{\includegraphics[width=5cm]{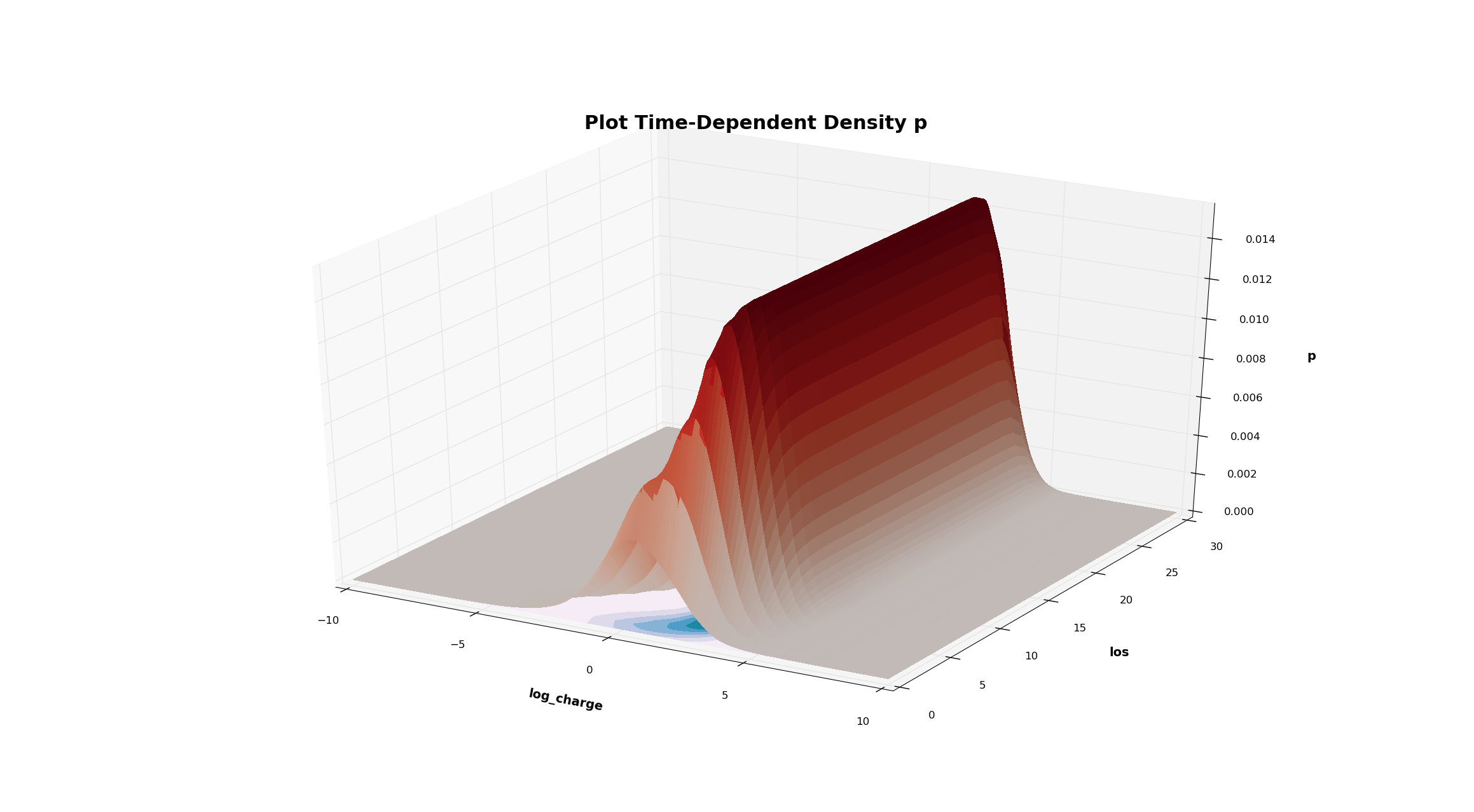}}
\caption{Function p (Prospective 1)}\label{fig:p_angle1}
\end{subfigure}
\hspace{1cm}
\begin{subfigure}[b]{5cm}
\centering
\frame{\includegraphics[width=5cm]{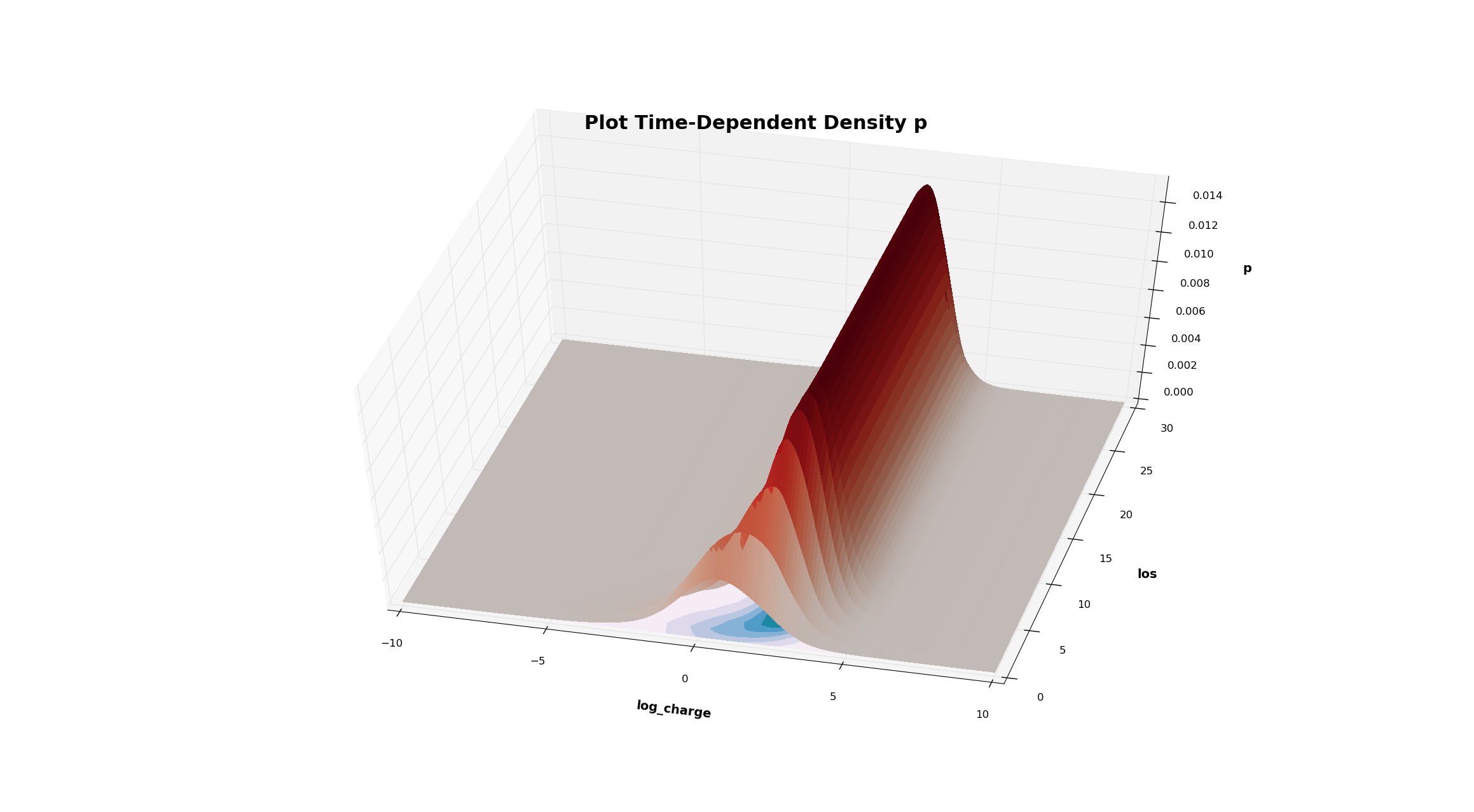}}
\caption{Function p (Prospective 2)}\label{fig:p_angle2}
\end{subfigure}
\end{center}
\end{minipage}

\begin{minipage}\linewidth
\begin{center}
\captionof{figure}{FML Estimation of Function $\tilde{q}_1$}\label{fig:q_tilde_1}
\includegraphics[width=10cm,height=5cm]{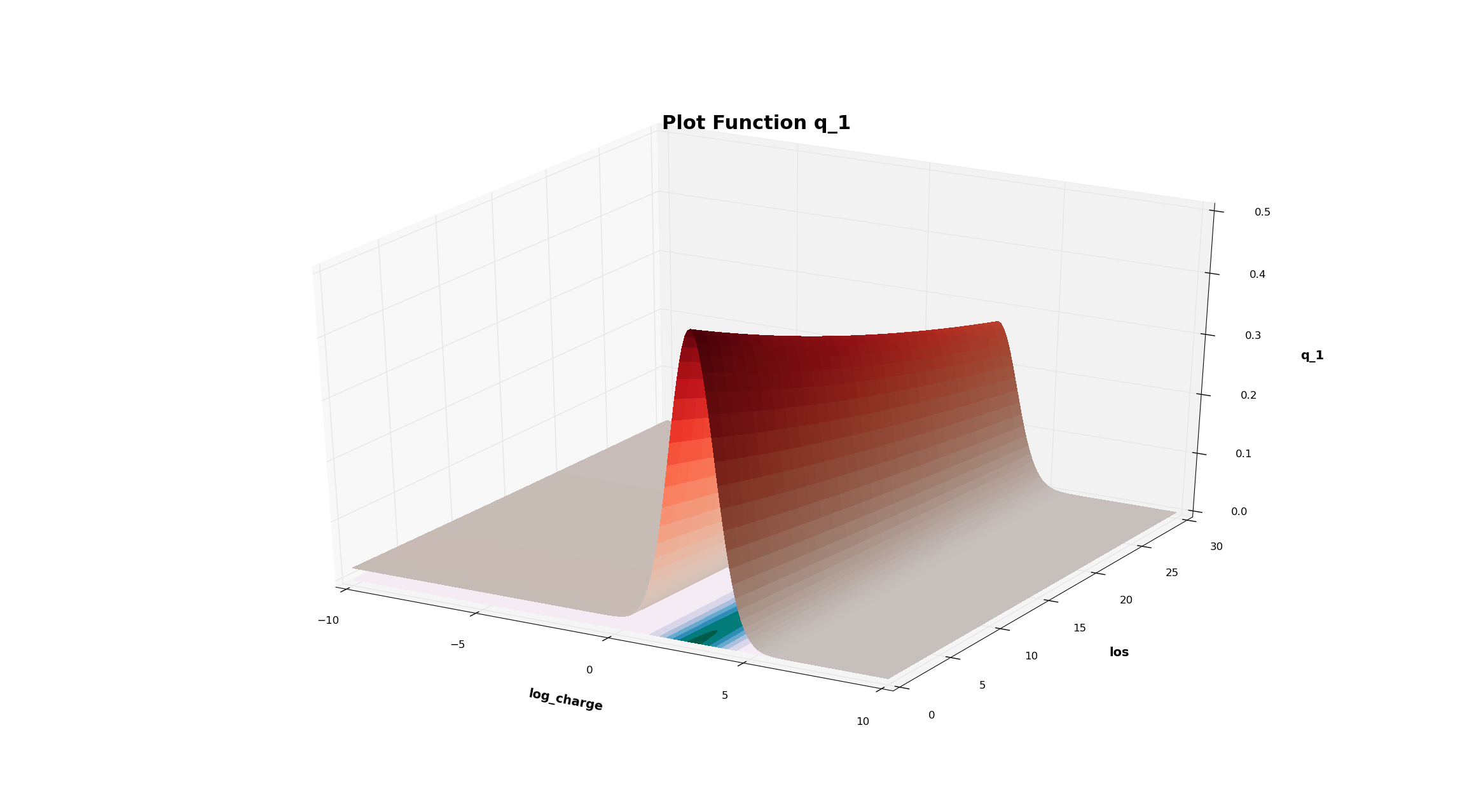}
\end{center}
\end{minipage}
The graph of function $p$ and $\tilde{q}_1$ is plotted in Figure \ref{fig:p_angle1}, \ref{fig:p_angle2} and \ref{fig:q_tilde_1}. From Figure \ref{fig:q_tilde_1}, it is easy to see that there exists an almost constant log-charge level for each fixed time $t$, below which patients has incentive to stay longer but above which patients are less likely to stay. We believe such a constant log-charge level corresponds to a psychological and/or medical threshold which plays a significant role in driving the treatment dynamics. From Figure \ref{fig:p_angle2}, there is a clear trend that as the length of time of staying increase, the mode of the accumulative charge goes up while the deviation from the mode gets shrink. Moreover, the increasing trend of mode and shrinking trend of deviation stablize when the length of time exceeds 10 days. 

It is worthwhile to mention that in \cite{gardiner2002longitudinal}, Gardiner  estimated the time-dependent expectation $E\left(Y_t\right)$ which can be derived from the function $p$ in the following way:
\begin{equation}
E\left(Y_t\right)=\int_{0}^{\infty}p\left(y,t\right)dy.
\end{equation} 
On the other hand, from the parametric family used in Gardiner \cite{gardiner2002longitudinal}, it is not possible to estimate the parametric form of $\tilde{q}_1$ function at all. Therefore, we believe our work extends \cite{gardiner2002longitudinal} and \cite{polverejan2003estimating} in the sense of providing more information to looking into the "black-box" of the treatment dynamics that patient experience.

\section{Discussion}

In this paper, we propose and parametrize the RGRST models and give a way to derive the joint probability density function of charge and LOS from RGRST models. We also show how the joint density function could help the application of FML method to estimate model parameters and resolve the endogeneity between charge and LOS.

There is an important open problem left, that is, the choice of the parametric form of RGRST models is not unique. As we emphasized in section 3.1, there could exist fairly different parametric families of RGRST models, each one of which could satisfy the three baseline principles \ref{criteria RGRST}, \ref{criteria data} and \ref{criteria mle}. Consequently, there is no way to uniquely determine a \enquote{best} parametric form. In fact, it turns out that different parametric forms have different types of implication to the treatment dynamics and the behavioral patterns of patient and/or doctors, and therefore, could serve for different analytic purposes. We believe that it is the analytic target that provides the ultimate criterion for the choice of parametric RGRST models.

In a series of related works, we will discuss a different way to parametrize the function $\tilde{q}_1$ and $p(.,0)$, from which we could convert a RGRST model to Coxian-Phase-Type models. An advantage of that parametrization is that we can estimate the \enquote{price} of each \enquote{phase} (in the associated Coxian-Phase-Type model) at each fixed time $t$. As suggested in \cite{marshall2007estimating}, a \enquote{phase} in a Coxian-Phase-Type model can be identified with a treatment stage that a patient may have to experience when stay in hospital, then the \enquote{price} information of each medical stage would be useful for analyzing patient's welfare and the dynamic management of medical resources. 
\begin{appendices}
\section{}
\subsection{Proof for Lemma \ref{lemma6}:}

Firstly, it is obvious that under the three condition in Lemma \ref{lemma6},
the process constructed by Equation \ref{q1_tilde construction} is indeed a RGRST process.

Next, consider the following partition of the event $\left\{ Y_{t}\in\left(y-\delta,y\right)\right\} $:

\begin{equation}
\label{A1}
\left\{ Y_{t}\in\left(y-\delta,y\right)\right\} =\left\{ Y_{t}\in\left(y-\delta,y\right),T\geq t\right\} \sqcup\left\{ Y_{t}\in\left(y-\delta,y\right),T<t\right\} 
\end{equation}

Obviously, 
\begin{equation}
\label{A2}
\begin{aligned}
&Prob\left\{ Y_{t}\in\left(y-\delta,y\right),T\geq t\right\} \\
=&Prob\left\{ \left(\omega,y_{0}\right):y_{0}\in\left(\tilde{g}\left(y-\delta,t,t\right),\tilde{g}\left(y,t,t\right)\right),\omega\leq\tilde{q}_{1}\left(\tilde{g}^{-1}\left(y_{0},0,t\right),t\right)\right\} \\
=&\int_{\tilde{g}\left(y-\delta,t,t\right)}^{\tilde{g}\left(y,t,t\right)}p\left(y_{0},0\right)\cdot\tilde{q}_{1}\left(\tilde{g}^{-1}\left(y_{0},0,t\right),t\right)dy_{0}
\end{aligned}
\end{equation}

and 

\begin{equation}
\label{A3}
\begin{aligned}
&Prob\left\{ Y_{t}\in\left(y-\delta,y\right),T<t\right\}   \\
= & Prob\left\{ \left(\omega,y_{0}\right):y_{0}\in\left(\tilde{g}\left(y,t,t\right),y\right),\tilde{q}_{1}\left(y,S\left(y_{0},y\right)\right)\leq\omega\leq\tilde{q}_{1}\left(y-\delta,S\left(y_{0},y-\delta\right)\right)\right\} \\
= & \int_{\tilde{g}\left(y,t,t\right)}^{y}p\left(y_{0},0\right)\cdot\left(\tilde{q}_{1}\left(y-\delta,S\left(y_{0},y-\delta\right)\right)-\tilde{q}_{1}\left(y,S\left(y_{0},y\right)\right)\right)dy_{0}
\end{aligned}
\end{equation}

where $S\left(y_{0},y\right):=\left\{ t:\tilde{g}\left(y,t,t\right)=y_{0}\right\} $.

Time dependent density induced by $Y_{t}$ satisfy:

\begin{equation}
\label{A4}
\begin{aligned}
&p\left(y,t\right)\\
= & \lim_{\delta\rightarrow0}\frac{Prob\left\{ Y_{t}\in\left(y-\delta,y\right)\right\} }{\delta}\\
=&\lim_{\delta\rightarrow0} 
\left(
\begin{gathered}
\frac{\int_{\tilde{g}\left(y-\delta,t,t\right)}^{\tilde{g}\left(y,t,t\right)}p\left(y_{0},0\right)\cdot\tilde{q}_{1}\left(\tilde{g}^{-1}\left(y_{0},0,t\right),t\right)dy_{0}}{\delta}\\
+\\
\frac{\int_{\tilde{g}\left(y,t,t\right)}^{y}p\left(y_{0},0\right)\cdot\left(\tilde{q}_{1}\left(y-\delta,S\left(y_{0},y-\delta\right)\right)-\tilde{q}_{1}\left(y,S\left(y_{0},y\right)\right)\right)dy_{0}}{\delta}
\end{gathered}
\right)
\end{aligned}
\end{equation}

It is not hard to show:

\begin{equation}
\label{A5}
\lim_{\delta\rightarrow0}\frac{\int_{\tilde{g}\left(y-\delta,t,t\right)}^{\tilde{g}\left(y,t,t\right)}p\left(y_{0},0\right)\cdot\tilde{q}_{1}\left(\tilde{g}^{-1}\left(y_{0},0,t\right),t\right)dy_{0}}{\delta}=p\left(\tilde{g}\left(y,t,t\right),0\right)\cdot\tilde{q}_{1}\left(y,t\right)\cdot\frac{\partial\tilde{g}\left(y,t,t\right)}{\partial y}
\end{equation}

and 

\begin{equation}
\label{A6}
\begin{aligned}
&\lim_{\delta\rightarrow0}\frac{\int_{\tilde{g}\left(y,t,t\right)}^{y}p\left(y_{0},0\right)\cdot\left(\tilde{q}_{1}\left(y-\delta,S\left(y_{0},y-\delta\right)\right)-\tilde{q}_{1}\left(y,S\left(y_{0},y\right)\right)\right)dy_{0}}{\delta}\\
=&\int_{\tilde{g}\left(y,t,t\right)}^{y}p\left(y_{0},0\right)\cdot\left(-\frac{\partial\tilde{q}_{1}\left(y,S\left(y_{0},y\right)\right)}{\partial y}-\frac{\partial\tilde{q}_{1}\left(y,S\left(y_{0},y\right)\right)}{\partial t}\cdot\frac{\partial S\left(y_{0},y\right)}{\partial y}\right)dy_{0}
\end{aligned}
\end{equation}

Through analyzing the solution trajectory of the ODE $y'=\tilde{q}\left(y,t\right)$,
it is easy to verify 

\[
\frac{\partial S\left(y_{0},y\right)}{\partial y}=\frac{1}{\tilde{q}\left(y,S\left(y_{0},y\right)\right)}
\]

\begin{eqnarray*}
S\left(\tilde{g}\left(y,t,t\right),y\right) & = & t
\end{eqnarray*}

and 

\[
\frac{\partial\tilde{g}\left(y,s,s\right)}{\partial y}=\frac{1}{\tilde{q}\left(y,s\right)}\cdot\frac{-\partial\tilde{g}\left(y,s,s\right)}{\partial s}
\]

Consequently, we have 

\begin{equation}
\label{A7}
\begin{aligned}
&\int_{\tilde{g}\left(y,t,t\right)}^{y}p\left(y_{0},0\right)\cdot\left(-\frac{\partial\tilde{q}_{1}\left(y,S\left(y_{0},y\right)\right)}{\partial y}-\frac{\partial\tilde{q}_{1}\left(y,S\left(y_{0},y\right)\right)}{\partial t}\cdot\frac{\partial S\left(y_{0},y\right)}{\partial y}\right)dy_{0}\\
= & \int_{0}^{t}p\left(\tilde{g}\left(y,s,s\right),0\right)\cdot\left(-\frac{\partial\tilde{q}_{1}\left(y,s\right)}{\partial y}-\frac{\partial\tilde{q}_{1}\left(y,s\right)}{\partial t}\cdot\frac{1}{\tilde{q}\left(y,s\right)}\right)\cdot\frac{-\partial\tilde{g}\left(y,s,s\right)}{\partial s}ds\\
= & \int_{0}^{t}p\left(\tilde{g}\left(y,s,s\right),0\right)\cdot\left(-\frac{\partial\tilde{q}_{1}\left(y,s\right)}{\partial y}\cdot\tilde{q}\left(y,s\right)-\frac{\partial\tilde{q}_{1}\left(y,s\right)}{\partial t}\cdot\right)\cdot\frac{1}{\tilde{q}\left(y,s\right)}\cdot\frac{-\partial\tilde{g}\left(y,s,s\right)}{\partial s}ds\\
= & \int_{0}^{t}p\left(\tilde{g}\left(y,s,s\right),0\right)\cdot\left(-\frac{\partial\tilde{q}_{1}\left(y,s\right)}{\partial y}\cdot\tilde{q}\left(y,s\right)-\frac{\partial\tilde{q}_{1}\left(y,s\right)}{\partial t}\cdot\right)\cdot\frac{\partial\tilde{g}\left(y,s,s\right)}{\partial y}ds
\end{aligned}
\end{equation}

Combine Equation \ref{A4}-\ref{A7}, we have 

\begin{equation}
\begin{aligned}
p\left(y,t\right)=&\tilde{q}_{1}\left(y,t\right)\cdot\frac{\partial\tilde{g}\left(y,t,t\right)}{\partial y}\cdot p\left(\tilde{g}
\left(y,t,t\right),0\right)\\
&+\int_{0}^{t}p\left(\tilde{g}\left(y,s,s\right),0\right)\cdot\left(-\frac{\partial\tilde{q}_{1}\left(y,s\right)}{\partial y}\cdot\tilde{q}\left(y,s\right)-\frac{\partial\tilde{q}_{1}\left(y,s\right)}{\partial t}\cdot\right)\cdot\frac{\partial\tilde{g}\left(y,s,s\right)}{\partial y}ds
\end{aligned} \nonumber
\end{equation}

Finally, by \ref{A3} and \ref{A7} the the definition of joint probability density function, it can be easily verified that the joint density function of total charge and LOS can be expressed as claimed in Equation \ref{joint density 2nd expression}.

\subsection{Proof for Lemma \ref{lemma5}}

Firstly, notice that Lemma \ref{lemma5} can be obtained from the following to
conditions:

1. The joint probability density $P\left(y,t\right)$ of total charge at discharge day, $X_T$, and the LOS, $T$, can be expressed in the same way as in Equation \ref{joint density 2nd expression};

2. The time-dependent density function $p\left(y,t\right)$ induced by $\left\{ X_{t}\right\} $
is of the same form as in Equation \ref{time dependent density 2};

3. The directional derivative of $\tilde{q}_{1}$ in the direction
given by $\tilde{q}$ is always non-negative, which guarantees the process $\left\{ Y_{t}\right\} $
constructed in Lemma \ref{lemma5} is a well defined RGRST process.

For condition 2, we have to adopt similar trick as in the proof of
Lemma \ref{lemma6}. Indeed, we need to compute $A_{y,t,\delta}:=Prob\left\{ X_{t}\in\left(y-\delta,y\right),T\geq t\right\} $
and $B_{y,t,\delta}:=Prob\left\{ X_{t}\in\left(y-\delta,y\right),T<t\right\} $ and show that the limit of their sum (as $\delta\rightarrow 0$) has exactly the same form as expressed in Equation \ref{time dependent density 2}.

By the definition of function $\tilde{q}$ and $\tilde{q}_{1}$ in Equation \ref{unique eq}, it is not hard to show $B_{y,t,\delta}=\int_{y-\delta}^{y+\delta}
\tilde{p}\left(x,t\right)\cdot
\tilde{q}_{1}\left(x,t\right)dx$ (may need repeatedly use Ito's Lemma and the construction in the proof of the existence theorem of solutions to an ordinary differential equation \cite{peacock1983two}).
So, it remains to verify:

\begin{equation} \label{A3_2}
A_{y,t,\delta}=\int_{y-\delta}^{y}\int_{0}^{t}-\left(\frac{\partial\tilde{q}_{1}\left(x,s\right)}{\partial y}\cdot\tilde{q}\left(x,s\right)+\frac{\partial\tilde{q}_{1}\left(x,s\right)}{\partial t}\right)\tilde{p}\left(x,s\right)dsdx
\end{equation}

Notice that 
\begin{equation}
\Scale[0.9]{
\begin{gathered}
\left\{ X_{t}\in\left(y-\delta,y\right),T\geq t\right\}= \left\{ G_{t}\in\left(y-\delta,y\right),T\geq t\right\} \\
=  \bigcap_{\Delta>0}\bigcup_{\left\{ s_{i}=i\cdot\Delta:i=0,\dots,n,n\cdot\Delta\leq t<\left(n+1\right)\cdot\Delta\right\} }\left\{ G_{s_{i}}\in\left(y-\delta,y\right),s_{i}\leq T<s_{i+1}\right\} \\
=  \bigcap_{\Delta>0}\bigcup_{\left\{ s_{i}=i\cdot\Delta:i=0,\dots,n,n\cdot\Delta\leq t<\left(n+1\right)\cdot\Delta\right\} }\left(\left\{ G_{s_{i}}\in\left(y-\delta,y\right),s_{i}\leq T\right\} -\left\{ G_{s_{i}}\in\left(y-\delta,y\right),s_{i+1}\leq T\right\} \right)
\end{gathered}}
\end{equation}

, therefore

\begin{equation} \label{A3_5}
\Scale[0.9]{
\begin{aligned}
A_{y,t,\delta} = & \lim_{\Delta\rightarrow0}\sum_{i=0}^{n_{t,\Delta}}\left(E\left(\mathbf{1}_{\left\{ G_{s_{i}}\in\left(y-\delta,y\right)\right\} }\cdot\mathbf{1}_{\left\{ s_{i}\leq T\right\} }\right)-E\left(\mathbf{1}_{\left\{ G_{s_{i}}\in\left(y-\delta,y\right)\right\} }\cdot\mathbf{1}_{\left\{ s_{i}+\Delta\leq T\right\} }\right)\right)\\
 = & \lim_{\Delta\rightarrow0}\sum_{i=0}^{n_{t,\Delta}}\left(E\left(\mathbf{1}_{\left\{ G_{s_{i}}\in\left(y-\delta,y\right)\right\} }\cdot E\left(s_{i}\leq T|G_{s_{i}}\right)\right)-E\left(\mathbf{1}_{\left\{ G_{s_{i}}\in\left(y-\delta,y\right)\right\} }\cdot E\left(s_{i}+\Delta\leq T|G_{s_{i}}\right)\right)\right)
\end{aligned}}
\end{equation}

Notice that 
\begin{equation} \label{A3_6}
\Scale[0.8]{
\begin{aligned}
E\left(s+\Delta\leq T|G_{s}=x\right) & = \int_{0}^{\infty}E\left(s+\Delta\leq T|G_{s+\Delta},G_{s}=x\right)dP\left(G_{s+\Delta}|G_{s}=x\right)\\
& = \int_{0}^{\infty}E\left(s+\Delta\leq T|G_{s+\Delta}=x+\int_{s}^{s+\Delta}\epsilon_{\tau}d\tau,G_{s}=x\right)dP\left(\int_{s}^{s+\Delta}\epsilon_{\tau}d\tau|G_{s}=x\right)。
\end{aligned}}
\end{equation}

In addition, using the right-continuity of decision process $I\left(.,X_{t},t\right)$,
$\int_{s}^{s+\Delta}\epsilon_{\tau}d\tau\rightarrow0\, a.s$ as $\Delta\rightarrow0$
and the continuity of $\tilde{q}_{1}$,
we have: 
\begin{equation} \label{A3_3}
\lim_{\Delta\rightarrow 0}E\left(s+\Delta\leq T|G_{s}=x\right) = \tilde{q}_{1}\left(x,s\right)
\end{equation}
and
\begin{equation} \label{A3_4}
\Scale[0.8]{
\lim_{\Delta\rightarrow0}\left(E\left(s+\Delta\leq T|G_{s}=x\right)-\int_{0}^{\infty}E\left(s+\Delta\leq T|G_{s+\Delta}=x+\int_{s}^{s+\Delta}\epsilon_{\tau}d\tau\right)dP\left(\int_{s}^{s+\Delta}\epsilon_{\tau}d\tau|G_{s}=x\right)\right)=0},
\end{equation}

where $dP\left(dG_{s+\Delta}|G_{s}=x\right)$ is the conditional
probability inuced by $G_{s+\Delta}$ given $G_{s}=x$ and $dP\left(\int_{s}^{s+\Delta}\epsilon_{\tau}d\tau|G_{s}=x\right)$
is the conditional probability induced by the increment $\int_{s}^{s+\Delta}\epsilon_{\tau}d\tau$
given $G_{s}=x$. Plug \ref{A3_3}, \ref{A3_4} and \ref{A3_6} into \ref{A3_5}, we have the following

\begin{equation}
\label{A3_conclusion}
\Scale[0.8]{
\begin{aligned}
&A_{y,t,\delta}\\  
= & \lim_{\Delta\rightarrow0}\sum_{i=0}^{n_{t,\Delta}}\left(
\begin{gathered}
\int_{y-\delta}^{y}E\left(s_{i}\leq T|G_{s_{i}}=x\right)\cdot\tilde{p}
\left(x,s_{i}\right)dx\\
-\\
\int_{y-\delta}^{y}\int_{0}^{\infty}E\left(s_{i}+\Delta\leq T|G_{s_{i}+\Delta}=x+\int_{s_{i}}^{s_{i}+\Delta}\epsilon_{\tau}d\tau\right)dP\left(\int_{s_{i}}^{s_{i}+\Delta}\epsilon_{\tau}d\tau|G_{s_{i}}=x\right)\cdot\tilde{p}
\left(x,s_{i}\right)dx
\end{gathered}
\right)\\
= & \lim_{\Delta\rightarrow0}\sum_{i=0}^{n_{t,\Delta}}\left(
\begin{gathered}
\int_{y-\delta}^{y}\tilde{q}_{1}
\left(x,s_{i}\right)\cdot\tilde{p}
\left(x,s_{i}\right)dx\\
-\\
\int_{y-\delta}^{y}\int_{0}^{\infty}\tilde{q}_{1}\left(x+\int_{s_{i}}^{s_{i}+\Delta}\epsilon_{\tau}d\tau,s_{i}+\Delta\right)dP\left(\int_{s_{i}}^{s_{i}+\Delta}\epsilon_{\tau}d\tau|G_{s_{i}}=x\right)\cdot\tilde{p}
\left(x,s_{i}\right)dx
\end{gathered}
\right)\\
= & -\lim_{\Delta\rightarrow0}\sum_{i=0}^{n_{t,\Delta}}\frac{\int_{y-\delta}^{y}\left(\int_{0}^{\infty}\tilde{q}_{1}\left(x+\int_{s_{i}}^{s_{i}+\Delta}\epsilon_{\tau}d\tau,s_{i}+\Delta\right)-\tilde{q}_{1}\left(x,s_{i}\right)\right)dP\left(\int_{s_{i}}^{s_{i}+\Delta}\epsilon_{\tau}d\tau|G_{s_{i}}=x\right)\tilde{p}\left(x,s_{i}\right)dx}{\Delta}\cdot\Delta\\
= & -\lim_{\Delta\rightarrow0}\sum_{i=0}^{n_{t,\Delta}}\frac{\int_{y-\delta}^{y}\int_{0}^{\infty}\left(\frac{\partial q_{1}}{\partial y}\cdot\int_{s}^{s+\Delta}\epsilon_{\tau}d\tau+\frac{\partial q_{1}}{\partial t}\cdot\Delta\right)\left(x,s\right)dP\left(\int_{s_{i}}^{s_{i}+\Delta}\epsilon_{\tau}d\tau|G_{s_{i}}=x\right)\tilde{p}\left(x,s_{i}\right)dx}{\Delta}\cdot\Delta\\
= & -\lim_{\Delta\rightarrow0}\sum_{i=0}^{n_{t,\Delta}}\frac{\int_{y-\delta}^{y}\left(\frac{\partial\tilde{q}_{1}\left(x,s_{i}\right)}{\partial y}\cdot\int_{0}^{\infty}\left(\int_{s_{i}}^{s_{i}+\Delta}\epsilon_{\tau}d\tau\right)dP\left(\int_{s_{i}}^{s_{i}+\Delta}\epsilon_{\tau}d\tau|G_{s_{i}}=x\right)+\frac{\partial\tilde{q}_{1}\left(x,s_{i}\right)}{\partial t}\cdot\Delta\right)\tilde{p}\left(x,s_{i}\right)dx}{\Delta}\cdot\Delta\\
= & -\lim_{\Delta\rightarrow0}\sum_{i=0}^{n_{t,\Delta}}\int_{y-\delta}^{y}\left(\frac{\partial\tilde{q}_{1}\left(x,s_{i}\right)}{\partial y}\cdot E\left(\frac{\int_{s_{i}}^{s_{i}+\Delta}\epsilon_{\tau}d\tau}{\Delta}|G_{s_{i}}=x\right)+\frac{\partial\tilde{q}_{1}\left(x,s_{i}\right)}{\partial t}\right)\tilde{p}\left(x,s_{i}\right)dx\cdot\Delta\\
= & -\lim_{\Delta\rightarrow0}\sum_{i=0}^{n_{t,\Delta}}\int_{y-\delta}^{y}\left(\frac{\partial\tilde{q}_{1}\left(x,s_{i}\right)}{\partial y}\cdot E\left(\epsilon_{s_{i}}|G_{s_{i}}=x\right)+\frac{\partial\tilde{q}_{1}\left(x,s_{i}\right)}{\partial t}\right)\tilde{p}\left(x,s_{i}\right)dx\cdot\Delta\\
= & -\lim_{\Delta\rightarrow0}\sum_{i=0}^{n_{t,\Delta}}\int_{y-\delta}^{y}\left(\frac{\partial\tilde{q}_{1}\left(x,s_{i}\right)}{\partial y}\cdot\tilde{q}\left(x,s_{i}\right)+\frac{\partial\tilde{q}_{1}\left(x,s_{i}\right)}{\partial t}\right)\tilde{p}\left(x,s_{i}\right)dx\cdot\Delta\\
 = & \int_{y-\delta}^{y}\int_{0}^{t}-\left(\frac{\partial\tilde{q}_{1}\left(x,s\right)}{\partial y}\cdot\tilde{q}\left(x,s\right)+\frac{\partial\tilde{q}_{1}\left(x,s\right)}{\partial t}\right)\tilde{p}\left(x,s\right)dsdx,
\end{aligned}}
\end{equation}

therefore equality \ref{A3_2} is verified. 

On the other hand, it is easy
to see from the construction of $A_{y,t,\delta}$ that the expression $
-\left(\frac{\partial q_{1}}{\partial y}\cdot\tilde{q}+\frac{\partial q_{1}}{\partial t}\right)\cdot\tilde{p}$
gives the joint probability density function of $X_{T}$ and $T$, which verifies the condition 1. Moreover, as a joint density of two random variables, the expression $
-\left(\frac{\partial q_{1}}{\partial y}\cdot\tilde{q}+\frac{\partial q_{1}}{\partial t}\right)\cdot\tilde{p}$ should always be
non-negative as long as $\left\{ X_{t}\right\} $ is a well-defined
RGRST process. Then, following from the positivity of $\tilde{p}$,
$\frac{\partial q_{1}}{\partial y}\cdot\tilde{q}+\frac{\partial q_{1}}{\partial t}$
is non-positive over its domain $\left(0,\infty\right)^{2}$. So condition 3 holds. This
completes proof for Lemma \ref{lemma5}.

\section{} 
\subsection{Proof for Theorem \ref{theorem flexibility}}\label{Proof for Theorem flex}
Let $\tilde{g}\left(y,t,s\right)$ and $\tilde{g}^{-1}\left(y,t,s\right)$
be the functions constructed from $\tilde{q}$ as in Lemma \ref{lemma5}, it is
easy to check that $\tilde{g}>0$ within $\left\{ \left(y,t\right):\, y\leq\tilde{g}^{-1}\left(c,0,t\right)\right\} $.
Thus, the function $h:=\frac{f}{\frac{\partial\tilde{g}}{\partial y}}$
is well-defined and non-negative in $\left\{ \left(y,t\right):\, y\leq\tilde{g}\left(c,0,t\right)\right\} $.
Denote

\[
p\left(y,0\right):=\int_{0}^{\infty}h\left(\tilde{g}^{-1}\left(y,0,t\right),t\right)dt,
\]
it is fairly easy to check $p\left(.,0\right)$ is a well defined
density function and if there exists a RGRST process with $\tilde{q}$
given as above and $f$ as its derived joint density function of charge-LOS,
its initial density must be expressed as above. In fact, suppose there
exist some RGRST process represented by the triple $\left(p'\left(.,0\right),\tilde{q},\tilde{q}_{1}\right)$
with derived joint density $f$, then using Lemma \ref{lemma6}, we have

\begin{equation*}
\begin{aligned}
&\int_{0}^{\infty}h\left(\tilde{g}^{-1}\left(y,0,t\right),t\right)dt \\ = & \int_{0}^{\infty}\frac{f\left(\tilde{g}^{-1}\left(y,0,t\right),t\right)}{\frac{\partial\tilde{g}}{\partial y}\left(\tilde{g}^{-1}\left(y,0,t\right),t,t\right)}dt\\
 = & \int_{0}^{\infty}p'\left(\tilde{g}\left(\tilde{g}^{-1}\left(y,0,t\right),t,t\right),0\right)\cdot\left(\frac{-\partial\tilde{q}_{1}}{\partial y}\cdot\tilde{q}-\frac{\partial\tilde{q}_{1}}{\partial t}\right)\left(\tilde{g}^{-1}\left(y,0,t\right),t\right)dt\\
 = & \int_{0}^{\infty}p'\left(y,0\right)\cdot\left(\frac{-\partial\tilde{q}_{1}}{\partial y}\cdot\tilde{q}-\frac{\partial\tilde{q}_{1}}{\partial t}\right)\left(\tilde{g}^{-1}\left(y,0,t\right),t\right)dt\\
 = & p'\left(y,0\right)\cdot\int_{0}^{\infty}\left(\frac{-\partial\tilde{q}_{1}}{\partial y}\cdot\tilde{q}-\frac{\partial\tilde{q}_{1}}{\partial t}\right)\left(\tilde{g}^{-1}\left(y,0,t\right),t\right)dt\\
 = & p'\left(y,0\right)\cdot\left(\tilde{q}_{1}\left(y,0\right)-\lim_{t\rightarrow\infty}\tilde{q}_{1}\left(\tilde{g}^{-1}\left(y,0,t\right),t\right)\right)\\
 = & p'\left(y,0\right)=p\left(y,0\right).
 \end{aligned}
\end{equation*}
Using the function $p\left(.,0\right)$ as constructed above, we
can define $h':=\frac{f}{\tilde{p}}$ where $\tilde{p}$ is constructed
from $p\left(.,0\right)$ and $\tilde{g}$ in the same way as in Lemma
\ref{lemma6}, obviously, $h'\geq0$. Construct an advection equation as below:

\[
\frac{\partial k}{\partial y}\cdot\tilde{q}+\frac{\partial k}{\partial t}=-h'
\]

It turns out a solution to the following boundary value problem:

\begin{eqnarray*}
\frac{\partial k}{\partial y}\cdot\tilde{q}+\frac{\partial k}{\partial t} & = & -h'\\
k\left(.,0\right) & \equiv & 1
\end{eqnarray*}
is the $\tilde{q}_{1}$ function required by the theorem. In fact,
given $k$ as a solution to above boundary value problem and suppose
there exist some RGRST process represented by the triple $\left(p\left(.,0\right),\tilde{q},\tilde{q}_{1}\right)$
with derived joint density $f$, then using Lemma \ref{lemma6}, we have

\begin{eqnarray*}
\frac{\partial k\left(y,t\right)}{\partial y}\cdot\tilde{q}\left(y,t\right)+\frac{\partial k\left(y,t\right)}{\partial t} & = & -h'\left(y,t\right)\\
 & = & \frac{f\left(y,t\right)}{\tilde{p}\left(y,t\right)}\\
 & = & \frac{\partial\tilde{q}_{1}\left(y,t\right)}{\partial y}\cdot\tilde{q}\left(y,t\right)+\frac{\partial\tilde{q}_{1}\left(y,t\right)}{\partial t}.
\end{eqnarray*}
Therefore, to finish proof for this theorem, it suffices to show:

(1). there exist a unique solution to above boundary value problem,
and

(2). the resulted solution $\tilde{q}_{1}$ satisfies property \ref{Property 1} - \ref{Property 3}
as stated in section \ref{sec 2.2}. (By Lemma \ref{lemma6}, this condition guarantees the
existence of a RGRST process as required)

Existence and uniqueness of solution to above boundary problem is
guaranteed by the characteristic method as discussed in \cite{evans2010partial} and an analytic
expression for this solution is as below:

\[
\tilde{q}_{1}\left(y,t\right):=1-\int_{0}^{t}h'\left(\tilde{g}^{-1}\left(\tilde{g}\left(y,t,t\right),0,s\right),s\right)ds.
\]
Among Property  \ref{Property 1} - \ref{Property 3}, the only thing unchecked is that $0<\tilde{q}_{1}\leq1$,
this is true because 
\begin{eqnarray*}
1-\tilde{q}_{1}\left(y,t\right) & = & \int_{0}^{t}h'\left(\tilde{g}^{-1}\left(\tilde{g}\left(y,t,t\right),0,s\right),s\right)ds\\
 & = & \int_{0}^{t}\frac{f\left(\tilde{g}^{-1}\left(\tilde{g}\left(y,t,t\right),0,s\right),s\right)}{\frac{\partial\tilde{g}}{\partial y}\left(\tilde{g}^{-1}\left(\tilde{g}\left(y,t,t\right),0,s\right),s,s\right)}\cdot\frac{1}{p\left(\tilde{g}\left(y,t,t\right),0\right)}ds\\
 & = & \frac{\int_{0}^{t}\frac{f\left(\tilde{g}^{-1}\left(\tilde{g}\left(y,t,t\right),0,s\right),s\right)}{\frac{\partial\tilde{g}}{\partial y}\left(\tilde{g}^{-1}\left(\tilde{g}\left(y,t,t\right),0,s\right),s,s\right)}ds}{\int_{0}^{\infty}\frac{f\left(\tilde{g}^{-1}\left(\tilde{g}\left(y,t,t\right),0,s\right),s\right)}{\frac{\partial\tilde{g}}{\partial y}\left(\tilde{g}^{-1}\left(\tilde{g}\left(y,t,t\right),0,s\right),s,s\right)}dts}
\end{eqnarray*}
\section{}
\subsection{Proof for Theorem \ref{theorem validation}}\label{Proof for Theorem validation}
It is easy to solve the $\tilde{g}\left(y,t,s\right)=y\cdot\exp\left(-a\cdot s\right)$
and $\tilde{g}^{-1}\left(y,t,s\right)=y\cdot\exp\left(a\cdot s\right)$
from the initial value problems stated in Lemma \ref{lemma5} associated with
$\tilde{q}\left(y,t\right)=a\cdot y$. Therefore $\tilde{p}\left(y,t\right)=\frac{2\cdot\exp\left(-a\cdot t\right)}{\pi\gamma\cdot\left(1+\left(\frac{y\cdot\exp\left(-a\cdot t\right)}{\gamma}\right)^{2}\right)}$
given the Cauchy initial density. Plugging in $\tilde{p}$, $\tilde{q}$  and $\tilde{q}_{1}$ (Equation \ref{q1_tilde form}) into Equation \ref{joint density 2nd expression}, the functional form of joint
density $p_{Y_{T},T}$ can be easily shown to be as given in Equation
\ref{joint density for estimation}.

The relation between Marginal LOS density $p_{T}$, marginal charge
density $p_{Y_{T}}$ and the joint density $p_{Y_{T},T}$ is given
through the following integral

\begin{eqnarray}
p_{Y_{T}}\left(y\right) & = & \int_{[0,\infty)}p_{Y_{T},T}\left(y,t\right)dt \label{marginal charge}\\
p_{T}\left(t\right) & = & \int_{[0,\infty)}p_{Y_{T},T}\left(y,t\right)dy \label{marginal LOS}
\end{eqnarray}

Using Equation \ref{marginal charge}, \ref{marginal LOS} and \ref{joint density for estimation}, it is easy to check that 

\begin{equation}
\label{step}
\begin{aligned}
&
\begin{split} 
p_{T}\left(t\right) - \frac{2}{\pi\gamma}\cdot\sum_{n=1}^{N}\theta_{n}\cdot\left(e_{1,d_{n}}\cdot e^{\left(S_{n}-a\right)\cdot t}\cdot1_{d_{n}}\cdot\int_{[0,\infty)}\frac{\exp\left(-\frac{\left(\ln\left(y\right)-\mu_{n}\right)^{2}}{2\sigma_{n}}\right)}{\sqrt{2\pi}\sigma_{n}}dy\cdot a \right.\\
-\left. e_{1,d_{n}}\cdot e^{\left(S_{n}-a\right)\cdot t}\cdot S_{n}\cdot1_{d_{n}}\cdot\int_{[0,\infty)}\left(1-\Phi\left(\frac{\ln\left(y\right)-\mu_{n}}{\sigma_{n}}\right)\right)dy\right)
\end{split}
&\\
=&\frac{-2\cdot\exp\left(-a\cdot t\right)}{\pi\gamma}\times&\\
&
\begin{split}
\sum_{n=1}^{N}\theta_{n}\cdot\left(e_{1,d_{n}}\cdot e^{S_{n}\cdot t}\cdot1_{d_{n}}\cdot\int_{[0,\infty)}\frac{\left(\frac{y\cdot\exp\left(-a\cdot t\right)}{\gamma}\right)^{2}}{\left(1+\left(\frac{y\cdot\exp\left(-a\cdot t\right)}{\gamma}\right)^{2}\right)}\cdot\frac{\exp\left(-\frac{\left(\ln\left(y\right)-\mu_{n}\right)^{2}}{2\sigma_{n}}\right)}{\sqrt{2\pi}\sigma_{n}}dy\cdot a\right.\\
-\left. e_{1,d_{n}}\cdot e^{S_{n}\cdot t}\cdot S_{n}\cdot1_{d_{n}}\cdot\int_{[0,\infty)}\frac{\left(\frac{y\cdot\exp\left(-a\cdot t\right)}{\gamma}\right)^{2}}{\left(1+\left(\frac{y\cdot\exp\left(-a\cdot t\right)}{\gamma}\right)^{2}\right)}\cdot\left(1-\Phi\left(\frac{\ln\left(y\right)-\mu_{n}}{\sigma_{n}}\right)\right)dy\right)
\end{split}&
\end{aligned}
\end{equation}

Obviously, the right hand side of Equation \ref{step} is bounded by $C\cdot\exp\left(-a\cdot t\right)$
where the constant $C$ can be chosen to be a number no less than $\frac{2}{\pi\gamma}\cdot\max_{n}\left(E_{n}\right)\cdot\left(a+\max_{n}\left(\max_{t}\left(e_{1,d_{n}}\cdot e^{S_{n}\cdot t}\cdot S_{n}\cdot1_{d_{n}}\right)\right)\right)$
where $E_{n}$ is the expectation of log-normal distribution specified
by $\left(\mu_{n},\sigma_{n}\right)$. The second term in the left
hand side \ref{step} can be expressed as 

\begin{equation} \label{appro marginal charge}
\frac{2}{\pi\gamma}\cdot\sum_{n=1}^{N}\theta_{n}\cdot E_{n}\cdot\left(e_{1,d_{n}}\cdot e^{\left(S_{n}-a\right)\cdot t}\cdot\left(a-S_{n}\right)\cdot1_{d_{n}}\right)
\end{equation}
where we use the relation that $\int_{0}^{\infty}f\left(y\right)\cdot ydy=\int_{0}^{\infty}\left(1-F\left(y\right)\right)dy$
with $f$ being a probability density function over $[0,\infty)$
and $F$ is its cumulative distribution function. Since \[e_{1,d_{n}}\cdot e^{\left(S_{n}-a\right)\cdot t}\cdot\left(a-S_{n}\right)\cdot1_{d_{n}}\]
is the density function of a Coxian Phase Type distribution associated
with transition matrix $S_{n}-a$, expression \ref{appro marginal charge} is just a linear
combination of Coxian Phase Type distributions with all coefficients
positive. Consequently, the resulting function is a density function
of another Coxian Phase Type distribution with transition matrix $S=\oplus_{n=1}^{N}S_{n}$.

In sum, we obtain the desired decomposition for the marginal LOS density
as a Coxian Phase Type density (generalized\footnote{
Notice that for this new Coxian Phase Type distribution, it may not
be a probability distribution because its total mass (=$\frac{2}{\pi\gamma}\sum_{n=1}^{N}\theta_{n}E_{n}$)
may not be 1. But it is always finite and positive, therefore it is
different from a probability measure only by multiplying a scalar. For the purpose
of approximating the shape of density function, this result is
good enough. 
}) and a residual term controlled
by $C\cdot\exp\left(-a\cdot t\right)$.

For charge density $p_{Y_{T}}$, to show it has the right tail asymptotically
equivalent to the right tail of some log-normal density function, it suffices
to show 
\begin{equation} \label{step1}
\lim_{y\rightarrow\infty}\frac{1-\Phi\left(\frac{\ln\left(y\right)-\mu_{n}}{\sigma_{n}}\right)}{\exp\left(-\frac{\left(\ln\left(y\right)-\mu'_{n}\right)^{2}}{2\sigma_{n}^{2}}\right)}\in\left(0,\infty\right)
\end{equation}
 
\begin{equation} \label{step2}
\lim_{y\rightarrow\infty}\frac{\int_{0}^{\infty}\frac{2}{\pi\gamma\cdot\left(1+\left(\frac{y\cdot\exp\left(-a\cdot t\right)}{\gamma}\right)^{2}\right)}\cdot e_{1,d_{n}}\cdot e^{\left(S_{n}-a\right)\cdot t}\cdot1_{d_{n}}dt}{\frac{1}{y^{2}}}\in\left(0,\infty\right)
\end{equation}
 and 
\begin{equation} \label{step3}
\lim_{y\rightarrow\infty}\frac{\int_{0}^{\infty}\frac{-2}{\pi\gamma\cdot\left(1+\left(\frac{y\cdot\exp\left(-a\cdot t\right)}{\gamma}\right)^{2}\right)}\cdot e_{1,d_{n}}\cdot e^{\left(S_{n}-a\right)\cdot t}\cdot S_{n}\cdot1_{d_{n}}dt}{\frac{1}{y^{2}}}\in\left(0,\infty\right)
\end{equation}
for every $n=1,\dots,N$. Since then $p_{Y_{T}}\left(y\right)\sim\sum_{n}\exp\left(-\frac{\left(\ln\left(y\right)-\mu'_{n}\right)^{2}}{2\sigma_{n}^{2}}\right)\cdot\frac{c_{n}}{y^{2}}\sim c\cdot\exp\left(-\left(\ln\left(y\right)-\mu\right)^{2}\right)/y$
as $y\rightarrow\infty$ for some properly chosen constants $c$,
$c_{n}$'s and $\mu$, $\mu'_{n}$'s. Equation \ref{step1}, \ref{step2} and \ref{step3} follows
easily from L'Hôpital's rule. This completes the proof.

\subsection{Proof for Proposition \ref{identification }}\label{Proof for Proposition id}
Using Theorem \ref{existence and uniqueness}, Lemma \ref{lemma6}, the joint density function specified as
in Proposition \ref{theorem validation} and the regression equation \ref{structural equation II}, it is trivial
to check that the joint density function of charge and LOS for the group
of patients $\left(x,x'\right)$ can be generated by a RGRST process
represented by the triple $\left(p_{x,x'}\left(.,0\right),\tilde{q}_{x,x'},\tilde{q}_{1,x,x'}\right)$
as given above. In fact, directly plugging in the triple $\left(p_{x,x'}\left(.,0\right),\tilde{q}_{x,x'},\tilde{q}_{1,x,x'}\right)$
into expression \ref{joint density for estimation} leads to the desired result.

To show the uniqueness, it suffices to show that there cannot exist
two set of parameters \[\left(\gamma,a,\left\{ \left(\mu_{n},\sigma_{n}\right)\right\} ,\left\{ \left(S_{n},d_{n}\right)\right\} ,\left\{ \theta_{n}\right\} ,N\right)\]
such that the derived joint density functions corresponding to the
two set of parameters are always equal, i.e.
\begin{equation}
\label{restriction 3.2.15}
p_{Y_{T},T}^{1}\left(y,t\right)\equiv p_{Y_{T},T}^{2}\left(y,t\right)
\end{equation}
For $i = 1$ or $2$:
\begin{equation}
\begin{aligned}
p_{Y_{T},T}^{i}\left(y,t\right)= &\frac{2}{\pi\gamma_{i}\cdot\left(1+\left(\frac{y\cdot\exp\left(-a_{2}\cdot t\right)}{\gamma_{i}}\right)^{2}\right)}\times\\
&\begin{split}
\sum_{n=1}^{N}\theta_{i,n}\cdot\left(e_{1,d_{n}}\cdot e^{\left(S_{n}-a_{i}\right)\cdot t}\cdot1_{d_{n}}\cdot\frac{\exp\left(-\frac{\left(\ln\left(y\right)-\mu_{n}\right)^{2}}{2\sigma_{n}}\right)}{\sqrt{2\pi}\sigma_{n}}\cdot a_{i}\right.\\
\left.-e_{1,d_{n}}\cdot e^{\left(S_{n}-a_{i}\right)\cdot t}\cdot S_{n}\cdot1_{d_{n}}\cdot\left(1-\Phi\left(\frac{\ln\left(y\right)-\mu_{n}}{\sigma_{n}}\right)\right)\right)
\end{split}
\end{aligned}
\end{equation}

Here, W.L.O.G. we assume $p_{Y_{T},T}^{1}$ and $p_{Y_{T},T}^{2}$
have the same number of summands $N$ and for each $n$, the $n$
summands are identical for $p_{Y_{T},T}^{1}$ and $p_{Y_{T},T}^{2}$,
because if this is not the case, say there exist some summand appearing
in $p_{Y_{T},T}^{1}$ but not in $p_{Y_{T},T}^{2}$, we can add the
same term to $p_{Y_{T},T}^{2}$ and set the corresponding $\theta_{2,n}$
to be zero: then nothing would be changed. 

The condition \ref{restriction 3.2.15} enforces that $p_{Y_{T},T}^{1}\left(0,0\right)=\frac{2}{\pi\gamma_{1}}=\frac{2}{\pi\gamma_{2}}=p_{Y_{T},T}^{2}\left(0,0\right)$,
which is equivalent to $\gamma_{1}=\gamma_{2}$, i.e., the initial
distributions must be identical. Similarly, condition \ref{restriction 3.2.15} enforces: 
\begin{equation}
\label{step4}
\begin{aligned}
&p_{Y_{T},T}^{1}\left(y,0\right) \\
=&\frac{2}{\pi\gamma\cdot\left(1+\left(\frac{y}{\gamma}\right)^{2}\right)}\sum_{n=1}^{N}\left(
\begin{aligned}
&\theta_{1,n}\cdot a_{1}\cdot\frac{\exp\left(-\frac{\left(\ln\left(y\right)-\mu_{n}\right)^{2}}{2\sigma_{n}}\right)}{\sqrt{2\pi}\sigma_{n}}-\\
&e_{1,d_{n}}\cdot S_{n}\cdot1_{d_{n}}\cdot\theta_{1,n}\cdot\left(1-\Phi\left(\frac{\ln\left(y\right)-\mu_{n}}{\sigma_{n}}\right)\right)
\end{aligned}
\right)\\
= &p_{Y_{T},T}^{2}\left(y,0\right)\\
=& \frac{2}{\pi\gamma\cdot\left(1+\left(\frac{y}{\gamma}\right)^{2}\right)}\sum_{n=1}^{N}\left(
\begin{aligned}
&\theta_{2,n}\cdot a_{2}\cdot\frac{\exp\left(-\frac{\left(\ln\left(y\right)-\mu_{n}\right)^{2}}{2\sigma_{n}}\right)}{\sqrt{2\pi}\sigma_{n}}-\\
&e_{1,d_{n}}\cdot S_{n}\cdot1_{d_{n}}\cdot\theta_{2,n}\cdot\left(1-\Phi\left(\frac{\ln\left(y\right)-\mu_{n}}{\sigma_{n}}\right)\right)
\end{aligned}
\right)
\end{aligned}
\end{equation}
Since a sequence of functions with the form of \[\Scale[0.9]{\left(\exp\left(-\frac{\left(\ln\left(.\right)-\mu_{1}\right)^{2}}{2\sigma_{1}}\right),1-\Phi\left(\frac{\ln\left(.\right)-\mu_{1}}{\sigma_{1}}\right),\dots,\exp\left(-\frac{\left(\ln\left(.\right)-\mu_{k}\right)^{2}}{2\sigma_{k}}\right),1-\Phi\left(\frac{\ln\left(.\right)-\mu_{k}}{\sigma_{k}}\right)\right)}\]
are independent if and only if there does not exist $i\not=j$ for $i,j\in\left\{ 1,\dots,k\right\} $
satisfying $\exp\left(-\frac{\left(\ln\left(y\right)-\mu_{i}\right)^{2}}{2\sigma_{i}}\right)\equiv\exp\left(-\frac{\left(\ln\left(y\right)-\mu_{j}\right)^{2}}{2\sigma_{j}}\right)$,
the equality \ref{step4} enforces 
\[
a_{1}\cdot s\theta_{1,i}=a_{2}\cdot s\theta_{2,i}
\]

where $s\theta_{j,i}=\sum_{n_{i}\in N_{i}}\theta_{j,n_{i}}$ such
that $j=1$ or $\,2$ and the set $N_{i}\subset\left\{ 1,\dots,N\right\} $
satisfies $n_{i},n_{i}'\in N_{i}$, $\exp\left(-\frac{\left(\ln\left(y\right)-\mu_{n_{i}}\right)^{2}}{2\sigma_{n_{i}}}\right)\equiv\exp\left(-\frac{\left(\ln\left(y\right)-\mu_{n_{i'}}\right)^{2}}{2\sigma_{n_{i'}}}\right)$.
If $N_{1}=\left\{ 1,\dots,N\right\} $, $s\theta_{1,1}=s\theta_{2,1}=1$
and therefore $a_{1}=a_{2}$. If $N_{1}\not=\left\{ 1,\dots,N\right\} $,
denote $N_{2}=\left\{ 1,\dots,N\right\} /N_{1}$. Then from equality
\ref{step4}, we have 
\begin{eqnarray*}
a_{1}\cdot\sum_{n_{2}\in N_{2}}\theta_{1,n_{2}} & = & a_{2}\cdot\sum_{n_{2}\in N_{2}}\theta_{2,n_{2}}\\
a_{1}\cdot s\theta_{1,i} & = & a_{2}\cdot s\theta_{2,i}\\
a_{j}\cdot\sum_{n_{2}\in N_{2}}\theta_{j,n_{2}} & = & a_{j}\cdot\left(1-s\theta_{j,1}\right)\, j=1,2
\end{eqnarray*}

The three equalities enforce $a_{1}=a_{2}$. In sum, $a_{1}=a_{2}$
always hold. Then, by the uniqueness part of Theorem \ref{theorem flexibility}, the two
RGRST processes derived from the two sets of parameters $\left(\gamma_{1},a_{1},\left\{ \left(\mu_{n},\sigma_{n}\right)\right\} ,\left\{ \left(S_{n},d_{n}\right)\right\} ,\left\{ \theta_{1,n}\right\} ,N\right)$
and $\left(\gamma_{2},a_{2},\left\{ \left(\mu_{n},\sigma_{n}\right)\right\} ,\left\{ \left(S_{n},d_{n}\right)\right\} ,\left\{ \theta_{2,n}\right\} ,N\right)$
are identical. This completes the proof of uniqueness.

\renewcommand{\theequation}{\Alph{chapter}.\arabic{section}.\arabic{equation}}
\section{Out-Sample Fitting Plots}

\begin{minipage}\linewidth
\begin{center}
\captionof{figure}{Out-Sample Marginal Fitting of Log-charge and LOS by RGRST Model (1:1)}\label{fig:marginal out-sample}
\includegraphics[width=9cm,height=4cm]{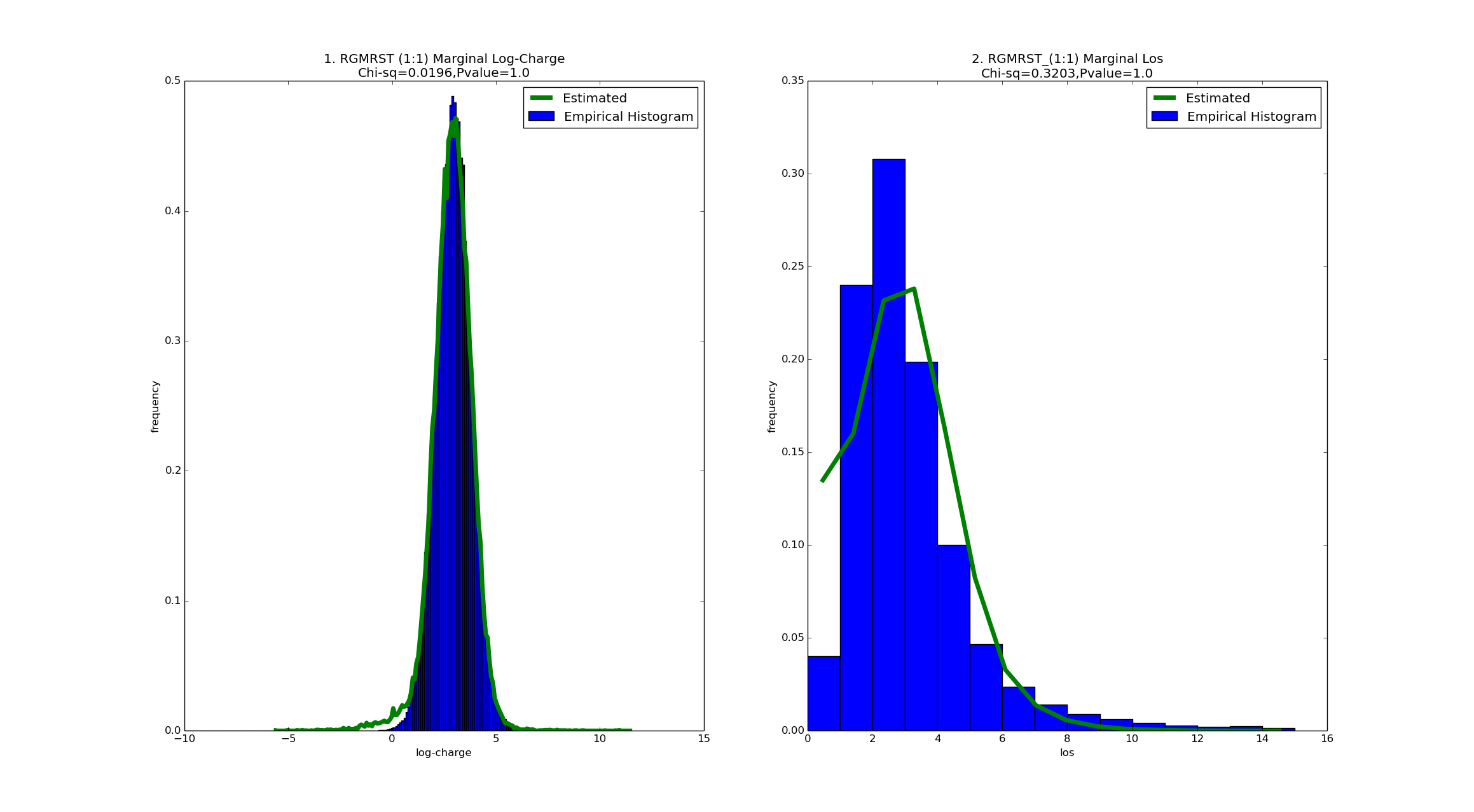}
\includegraphics[width=9cm,height=4cm]{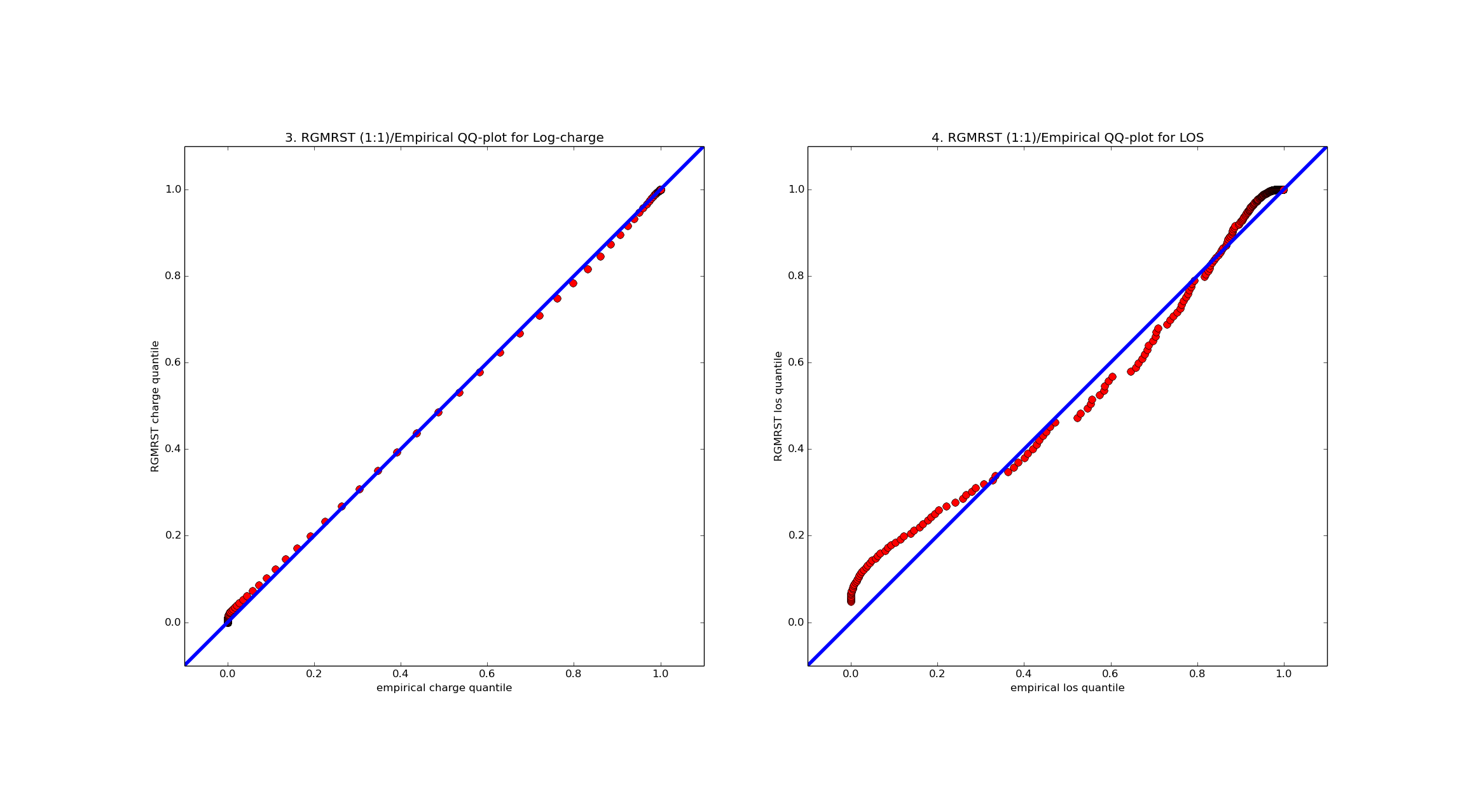}
\includegraphics[width=9cm,height=4cm]{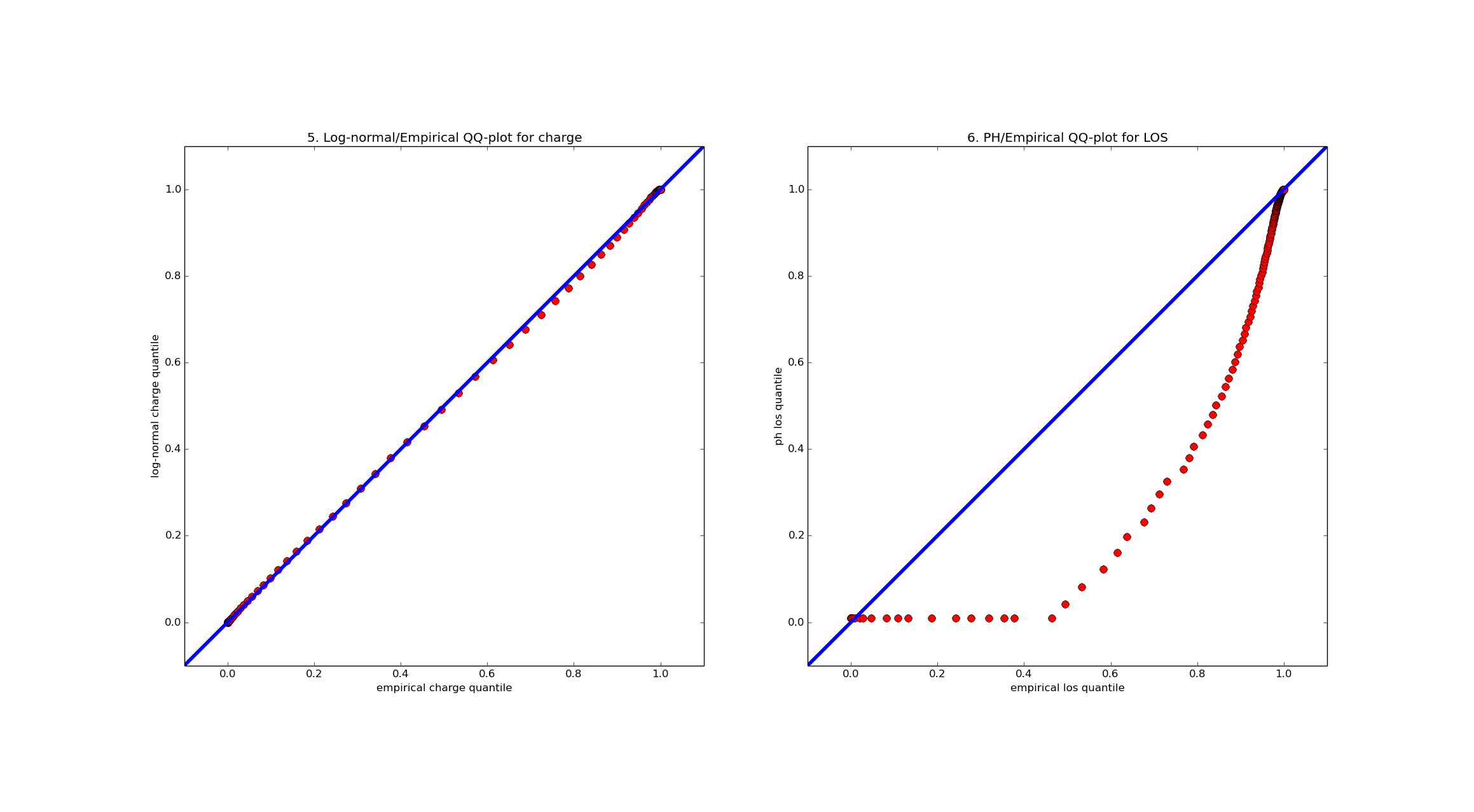}
\end{center}
\end{minipage}
\begin{minipage}\linewidth
\begin{center}
\captionof{figure}{Out-Sample Joint Fitting to Log-charge and LOS by RGRST (1:1)}\label{fig:joint out}
\includegraphics[width=14cm,height=7cm]{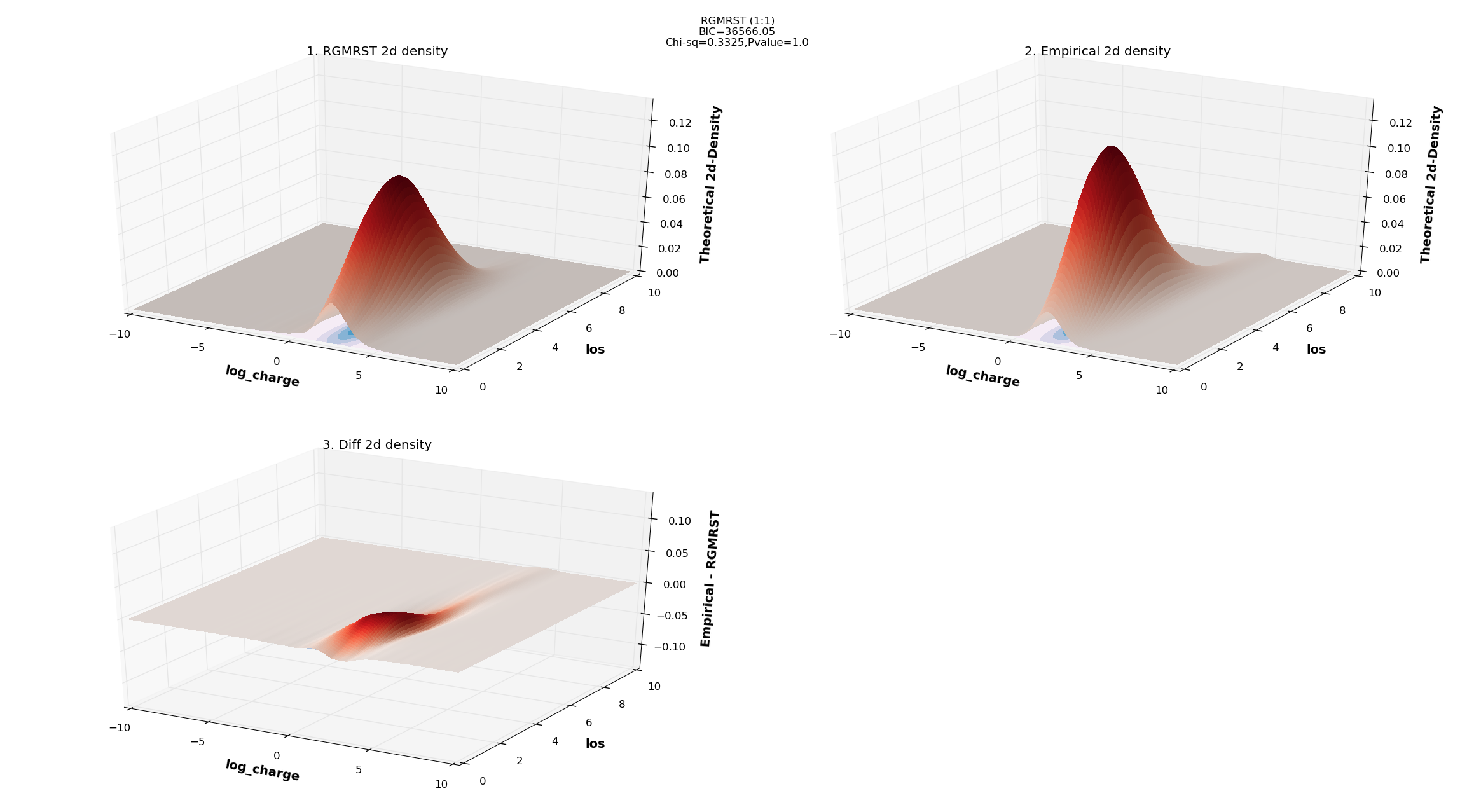}
\begin{minipage}{1\textwidth}
{\footnotesize Plot 1 is the joint density of log-charge and LOS derived from RGRST (1:1). Plot 2 is the empirical joint density obtained from Gaussian kernel density estimation (KDE) with kernel width 0.15 for log-charge and 1 for LOS. Plot 3 is obtained from subtracting Plot 1 from Plot 2.\par}
\end{minipage}
\end{center}
\end{minipage}

\end{appendices}

%
%

%

\renewcommand{\footnotesize}{\fontsize{8pt}{10pt}\selectfont}

\end{document}